\documentclass[journal=jacsat,manuscript=article]{achemso}

\usepackage[version=3]{mhchem} % Formula subscripts using \ce{}

\usepackage[version=3]{mhchem} % Formula subscripts using \ce{}
\usepackage{physics}
\usepackage{amsmath, amssymb}
\usepackage{parskip} 
\usepackage[utf8]{inputenc}
\usepackage{subcaption}
\usepackage[toc,page]{appendix}
\usepackage{algorithm}
\usepackage{xcolor}
\usepackage{algpseudocode}
 \SectionNumbersOn

\author{Rishabh Gupta}
\affiliation[Purdue University]
{Department of Chemistry, Purdue University, West Lafayette, IN, USA}
\author{Raja Selvarajan}
\affiliation[Purdue University]
{Department of Physics and Astronomy, Purdue University, West Lafayette, IN, USA}

\author{Manas Sajjan}
\affiliation[Purdue University]
{Department of Chemistry, Purdue University, West Lafayette, IN, USA}

\author{Raphael D. Levine}
\affiliation[Hebrew University]
{The Fritz Haber Center for Molecular Dynamics and Institute of Chemistry, The Hebrew University of Jerusalem, Jerusalem 91904, Israel
}
\alsoaffiliation[University of California]
{Department of Chemistry and Biochemistry and  Department of Molecular and
Medical Pharmacology, David Geffen School of Medicine, University of California, Los
Angeles, CA 90095, USA}

\author{Sabre Kais}
\email{kais@purdue.edu}
\affiliation[Purdue University]
{Department of Chemistry, Department of Physics and Astronomy, and Purdue Quantum Science and Engineering Institute, Purdue University, West Lafayette, IN, USA}

\title{Hamiltonian learning from time dynamics using variational algorithms}
\begin{document}
\maketitle

\begin{abstract}
The Hamiltonian of a quantum system governs the dynamics of the system via the Schrodinger equation. In this paper, the Hamiltonian is reconstructed in the Pauli basis using measurables on random states forming a time series dataset. The time propagation is implemented through Trotterization and optimized variationally with gradients computed on the quantum circuit. We validate our output by reproducing the dynamics of unseen observables on a randomly chosen state not used for the optimization. Unlike existing techniques that try and exploit the structure/properties of the Hamiltonian, our scheme is general and provides freedom with regard to what observables or initial states can be used while still remaining efficient with regard to implementation. We extend our protocol to doing quantum state learning where we solve the reverse problem of doing state learning given time series data of observables generated against several Hamiltonian dynamics. We show results on Hamiltonians involving $XX$, $ZZ$ couplings along with transverse field Ising Hamiltonians and propose an analytical method for the learning of Hamiltonians consisting of generators of the $SU(3)$ group. This paper is likely to pave the way toward using Hamiltonian learning for time series prediction within the context of quantum machine learning algorithms.

\end{abstract}

\section{Introduction}

Quantum tomography \cite{nielson, kais, qubits, cramer, paris2004quantum, artiles2005invitation} includes the study of quantum systems (state tomography) and the dynamics (process tomography) that govern these systems. While in process tomography, we are interested in reconstructing the Hamiltonian that governs the dynamics of the system \cite{yu2022practical, haah2021optimal, krastanov2019stochastic, evans2019scalable, bairey2019learning, qi2019determining}, in quantum state tomography we characterize the quantum mechanical state by the measurement of expectation values of an informationally complete (IC) set of Hermitian operators given multiple copies of a system. Within process tomography, we have direct methods that include standard quantum process tomography \cite{PhysRevLett.78.390}, ancilla assisted process \cite{PhysRevLett.86.4195, PhysRevLett.90.193601, PhysRevLett.91.047902} and indirect methods that make use of system dynamics, quantum entanglement, and average fidelity information \cite{PhysRevLett.88.217901, PhysRevLett.89.127902, PhysRevLett.95.240407}.  Some of the recent work in state tomography includes using permutation invariance \cite{PhysRevLett.105.250403}, entropy maximization \cite{gupta2021convergence, gupta2021maximal, gupta2022variation}, shadow tomography \cite{aransonst} , incoherent measurements \cite{chen2022tight} and ensemble averaging \cite{huang2020predicting}. 

In principle, the Hamiltonian learning  problem requires estimating a number of parameters that increases exponentially with the size of the system. But in reality, most physical Hamiltonians can be described by only a few-body relevant interaction terms that scale polynomially with the size of the quantum systems. While the conventional Hamiltonian learning techniques require the preparation of ground or thermal states   which is still a challenge \cite{schuch2008computational, bilgin2010preparing}, our Hamiltonian learning algorithm entirely circumvents this problem by providing independence on the choice of the quantum states as well as on the observables whose time dynamics can be easily recorded. For this work, we focus on \textit{n}-qubit Hamiltonians involving at most two-body interactions and can be defined as $H = \sum_{i,j>i}^n c_{ij}\sigma_i\sigma_j$ where $\sigma's$ are the Pauli string operators ($\sigma_x$, $\sigma_y$, $\sigma_z$, $\sigma_i$). To reconstruct the Hamiltonian of such a system, it is sufficient to estimate the parameters $\{c_{ij}\}$ and one of the ways of doing so is through optimization using variational algorithms. Here, we do a study of Hamiltonian learning using the time dynamics of observables measured on different quantum states using variational algorithms run on parameterized quantum circuits (PQC's). The coefficients of the Hamiltonian are reconstructed in the Pauli basis through an optimization scheme solved on PQC's. Variational quantum algorithms (VQA) \cite{mcclean2016theory} are seen as one of the promising techniques that can likely explore the power of computation on quantum circuits \cite{bravo2019variational, huang2019near, larose2019variational, cerezo2020variational, peruzzo2014variational, chen2021variational}. Some of the applications include electronic structure calculations \cite{sureshbabu2021implementation,sajjan2021quantum}, explore low energy symmetry states \cite{selvarajan2022variational}, and supervised machine learning \cite{dixit2021training}. The interested reader may refer to \cite{D2CS00203E} for a thorough study of variational algorithms and their use in quantum machine learning for chemistry and physics. 

We show how the learning of the Hamiltonian is affected by increasing the number of observables, quantum states, and the sampling frequency that we choose over the time interval. We exploit the fact that given sufficient observables to characterize the dynamics of a Hamiltonian, we can uniquely converge on reconstructing the exact Hamiltonian. Further, we show examples that exploit the knowledge of the Hamiltonian locality and try reconstructing the Hamiltonian of the system with very high fidelity. In practice, one does not have access to the actual Hamiltonian. We thus create validation schemes that can be used to inform us about the statistical knowledge required for reconstructing the Hamiltonian. 

Following this, we extend the scope of the method to state learning where the tools developed straightforwardly extend with a hardware efficient ansatz used to model the state. Unlike conventional quantum state learning of providing a state description using tomography we here would like to characterize the state as parameters of an ansatz that maps to it. The methods used here are found to be robust to random parameter Hamiltonians and initial starting states used.

\section{Methodology} \label{method}

We describe the problem in the context of spin systems as they directly map to the space of qubits over which the computations shall be performed. Let an unknown Hamiltonian act on a set of $n$ qubits whose initial state we get to choose/prepare. The experimenter measures the time dynamics of random observables chosen on this system. We would like to be able to predict the Hamiltonian that closely approximates the dynamics of the observables. Throughout our discussion, we shall restrict ourselves to second-order Hamiltonian couplings. Generalization of this can be extended to higher-order couplings. In the absence of any knowledge of the system's Hamiltonian, we consider a generic Hamiltonian with all to all coupling. Thus,

\begin{equation}
    H = \sum_{i,j,\beta,\gamma} \; J_{ij\beta\gamma}\sigma^{\beta}_i \sigma^{\gamma}_j
\end{equation}

where $\beta, \gamma$ indexes the Pauli matrices $\{X,Y,Z\}$. Let $\{\ket{\psi_i}\}$ be the set of states over which the Hamiltonian is allowed to act and let $\{O_{\alpha}\}$ be the observables whose dynamics have been recorded/sampled at discrete intervals of time. We work in the framework where $\{\ket{\psi_i}\}$ can be efficiently prepared and $\{O_{\alpha}\}$ can be efficiently measured. Let $O_{\alpha}(k \Delta t)$ refer to the measurement of observable $O_{\alpha}$ made at the timestep $k$, where $\Delta t$ refers to the time interval between the measurements. 

To start with we randomly initialize the parameters of the $J_{ij}$ of the Hamiltonian. We then construct the unitary operator $U(t) = exp(-iHt)$ that generates the time dynamics of $H$ using Trotterization \cite{cao2019quantum}. The constructed unitary $U$ is allowed to act on each of the starting states $\ket{\psi_i}$ and observables $\ket{O_{\alpha}}$ is measured at intervals of $\Delta t$ for a total of $N$ timesteps, thus running for the evolution for the duration of $N\Delta t$. Having a Trotter implementation of $U$ for time $\Delta t$ allows us to trivially extend the dynamics for time $k\Delta t$ by repeating $k$ layers of our base implementation. Just like any other machine learning algorithm, we have a scheme that implements the dynamics of a generic Hamiltonian $H$ and data points $O_{\alpha}(k \Delta t)$ measured. We are now left with minimizing a cost function of our choice. We choose the 2-norm function to minimize between $O_{\alpha}(k \Delta t)$ and $O^{obs}_{\alpha,i}(k \Delta t)$ over the states $\ket{\psi_i}$. Thus our cost function is,

\begin{equation}
    \mathrm{Cost} = \sum_{\alpha ,i,k} (\bra{\psi_i}U^{\dagger}(k\Delta t) \; O_{\alpha} \; U(k\Delta t)\ket{\psi_i} - O^{obs}_{\alpha,i,k}) ^2
\end{equation}

where the summation $\alpha$ indexes the summation over observables, $i$ indexes the summation over various starting states over which the dynamics is carried, and $k$ indicates the time steps over the evolution. Here $O^{obs}_{\alpha,i,k}$ refers to the observable $O^{obs}$ measured on $\ket{\psi_i}$ at the timestep $k$. To minimize the cost function we use standard gradient descent (SGD) to compute the gradients $\frac{d \mathrm{Cost}}{dJ_{ij}}$. Within Trotterization one can trivially map the couplings $J_{ij}$ directly to the circuit parameters of the constructed $U(\Delta t)$. This allows us to trivially compute the gradient to circuit parameters using parameter shift rule \cite{crooks2019gradients}. Alternatively, the analytical expression for gradients can be directly computed on a quantum circuit for these specific Trotter circuits as shown in Appendix \ref{gradient}\cite{selvarajan2021prime}.

As a measure of validation, the experimenter then validates against a new observable not yet measured on the system. If the Hamiltonian fails to reproduce an approximate comparison, the experimenter chooses to either increase the number of starting states or the number of observables over which the dynamics are observed. We show a shallow-depth implementation of Trotterization for an XYZ Hamiltonian in Appendix \ref{trotter}.

\subsection*{Extension to State learning}

We showed how using multiple starting states one can learn the Hamiltonian given the dynamics of the observables. We now solve the inverse problem. We use multiple Hamiltonians to learn/prepare a state of interest. Thus formally,  let $\{H_i\}$ be the set of Hamiltonians over which the state is allowed to evolve and let $\{O_{\alpha}\}$ be the observables whose dynamics have been recorded/sampled at discrete intervals of time. We work in the framework where $\{U_i = exp(-i H_it)\}$ can be efficiently synthesized and $\{O_{\alpha}\}$ can be efficiently measured. Let $O_{\alpha}(k \Delta t)$ refer to the measurement of observable $O_{\alpha}$ made at timestep $k$, where $\Delta t$ refers to the time interval of each timestep. 

\begin{figure}[ht!]
  \centering 
\includegraphics[width=3.2in]{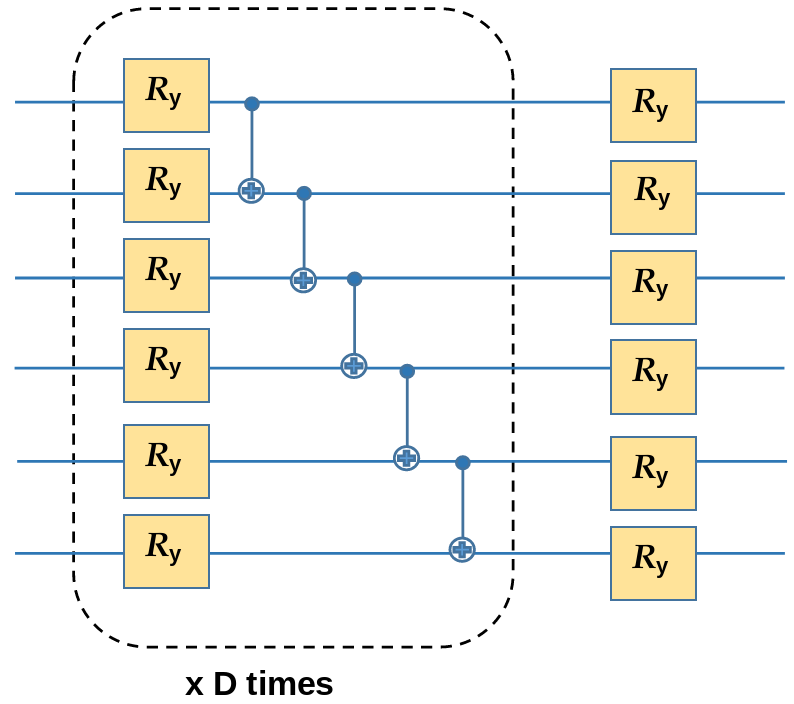} 
\caption{Ansatz $V(\vec{\beta})$ used to generate the state $\ket{\psi(\vec{\beta})}$ with real coefficients. The ansatz is made of $R_y$ Pauli rotation gates to restrict to real coefficients and a ladder of CNOT gates that can be used to create the entanglement. }
\label{ansatz}
\end{figure} 

We choose an ansatz $V(\vec{\beta})$ to prepare the state of interest $\ket{\psi(\vec{\beta})}$. Fig \ref{ansatz} shows a schematic representation of the ansatz used to create a state with real coefficients.  Thus we have,  $V(\vec{\beta})\ket{0}^{\bigotimes n} = \ket{\psi(\vec{\beta})}$, where $n$ is the number of qubits used to represent the state. We then impose the constraints of the dynamics recorded by $O_{\alpha}$. Thus we end up with an optimization problem with the following cost function to be minimized over the variables $\vec{\beta}$,

\begin{equation}
    \mathrm{Cost} = \sum_{\alpha ,i,k} (\bra{\psi(\vec{\beta})}U_i^{\dagger}(k\Delta t) \; O_{\alpha} \; U_i(k\Delta t)\ket{\psi(\vec{\beta})} - O^{obs}_{\alpha,i,k}) ^2 \label{cost1}
\end{equation}

where the summation $\alpha$ indexes the summation over observables, $i$ indexes the summation over various unitaries $U_i = exp(-i H_it)$ that implements the dynamics, and $k$ indicates the time steps over the evolution. Here $O^{obs}_{\alpha,i,k}$ refers to the observable $O^{obs}$ measured on $\ket{\psi_i}$ at the timestep $k$. To minimize the cost function we use SGD to compute the gradients $\frac{d \mathrm{Cost}}{d\beta_{s}}$, where $s$ labels the indices of $\vec{\beta}$. These gradients can be computed again trivially using the parameter shift rule. 

\section{Results and discussion} \label{result}
In this current work, we propose a variational approach for Hamiltonian and quantum state learning based on the time dynamics of various observables. We considered various quantum systems and performed numerical simulations in IBM's Qiskit \cite{Qiskit} to test and validate the proposed formalism.

\subsection{2-local Hamiltonian with all $\sigma_z\sigma_z$ and $\sigma_x\sigma_x$ coupling}
\begin{figure}[ht!]
  \centering 
\includegraphics[width=3.2in]{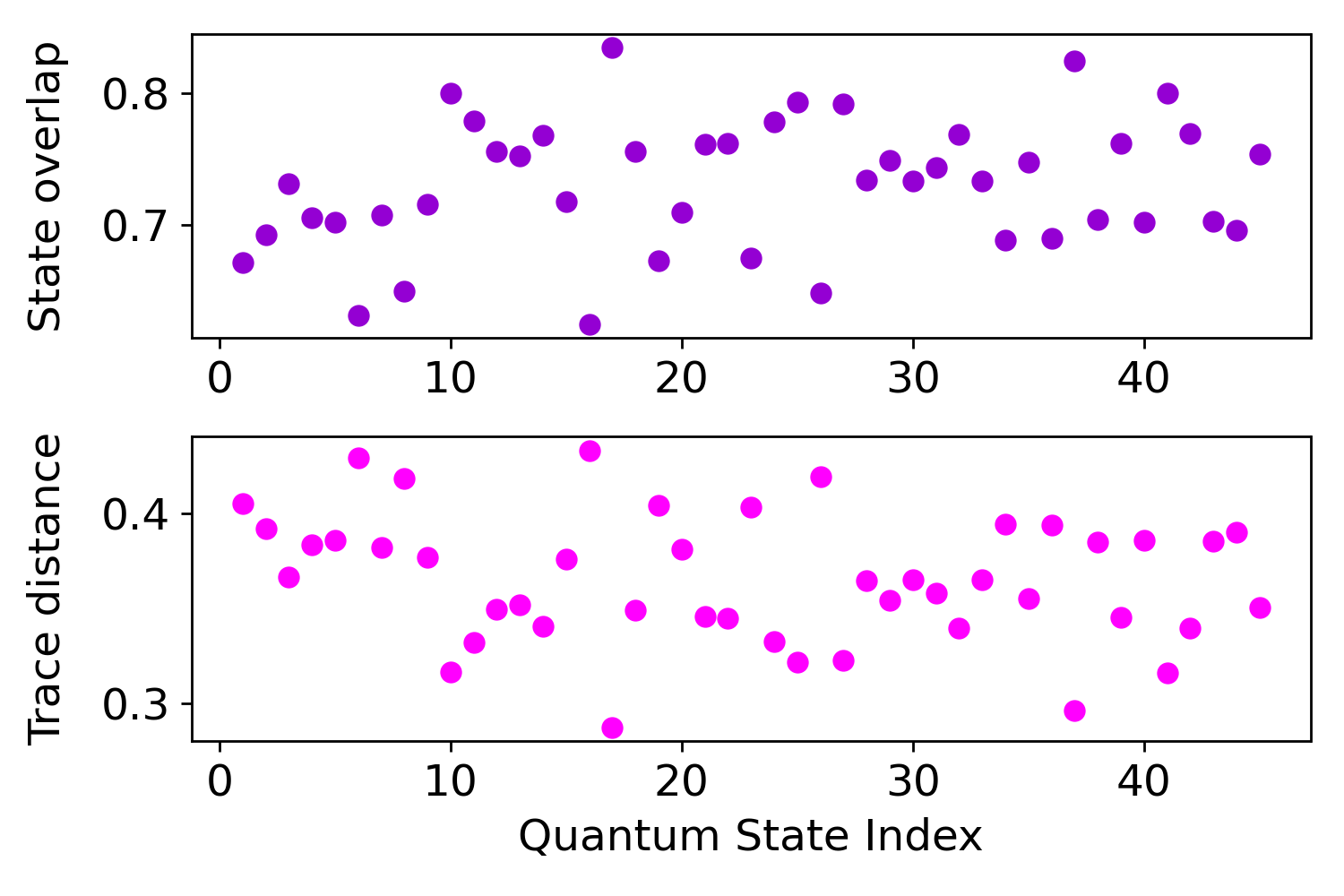} 
\caption{State overlap and trace distance between the considered quantum states to depict that the proposed protocol is independent of the choice of the starting quantum states.}
\label{fig_rand_states}
\end{figure}

To test the validation of our proposed approach for Hamiltonian tomography, we first considered a general 2-local Hamiltonian of the form: $H = \sum_{i,j>i}c_{ij}\sigma_z^i\sigma_z^j + \sum_{i,j>i}d_{ij}\sigma_x^i\sigma_x^j$ where $(c_{ij}, d_{ij})$  are randomly chosen coupling parameters between qubit-\textit{i} and qubit-\textit{j}. This Hamiltonian H is the true Hamiltonian that we want to learn and thereby predict the time dynamics of certain unknown observables. The cost function of the optimization process is defined as the mean square error between the true and reconstructed time series expectation values of the different observables considered computed using random starting quantum states. To verify that the starting quantum states are chosen at random we plot in Figure \ref{fig_rand_states} the state overlap and the trace distance, T($\rho,\sigma$) = $\frac{1}{2}||\rho-\sigma||_1$, between the considered quantum states. The observables correspond to the 2-point correlation functions: $\sigma_z^i\sigma_z^j$ and $\sigma_x^i\sigma_x^j$ whose time dynamics can be obtained from experiments. We, hereby, present an extensive list of results and inferences from the learning of the above Hamiltonian.

One of the measures that we considered to test the performance of our approach is the trace distance between the true and the reconstructed Hamiltonian. The trace distance in the case of two Hamiltonians is defined as: if $U_H = \exp\{-iHt\}$ and $U_K = \exp\{-iKt\}$ are the evolution operators for the true (\textit{H}) and the learning (\textit{K}) Hamiltonians then the trace distance is $T(H,K) = ||U_H(t)^\dagger U_K(t) - I||_2$. As described in Section \ref{method}, the different variables in our approach are the number of time steps ($N_T$) up to which the expectation values of the different observables need to be recorded, the number of starting quantum states ($N_S$), and the number of observables ($N_O$) required for accurate learning of the Hamiltonian. Since in an experiment, not all observables can be measured with high fidelity so we try to use a limited number of observables for which the time dynamics are recorded for various starting states. We vary each of these variables and calculate the trace distance in each case to find an optimized set of ($N_T$,$N_S$,$N_O$) as per the requirement of the experiment. \\
\begin{figure}[ht!]
  \centering 
\includegraphics[width=5.5in]{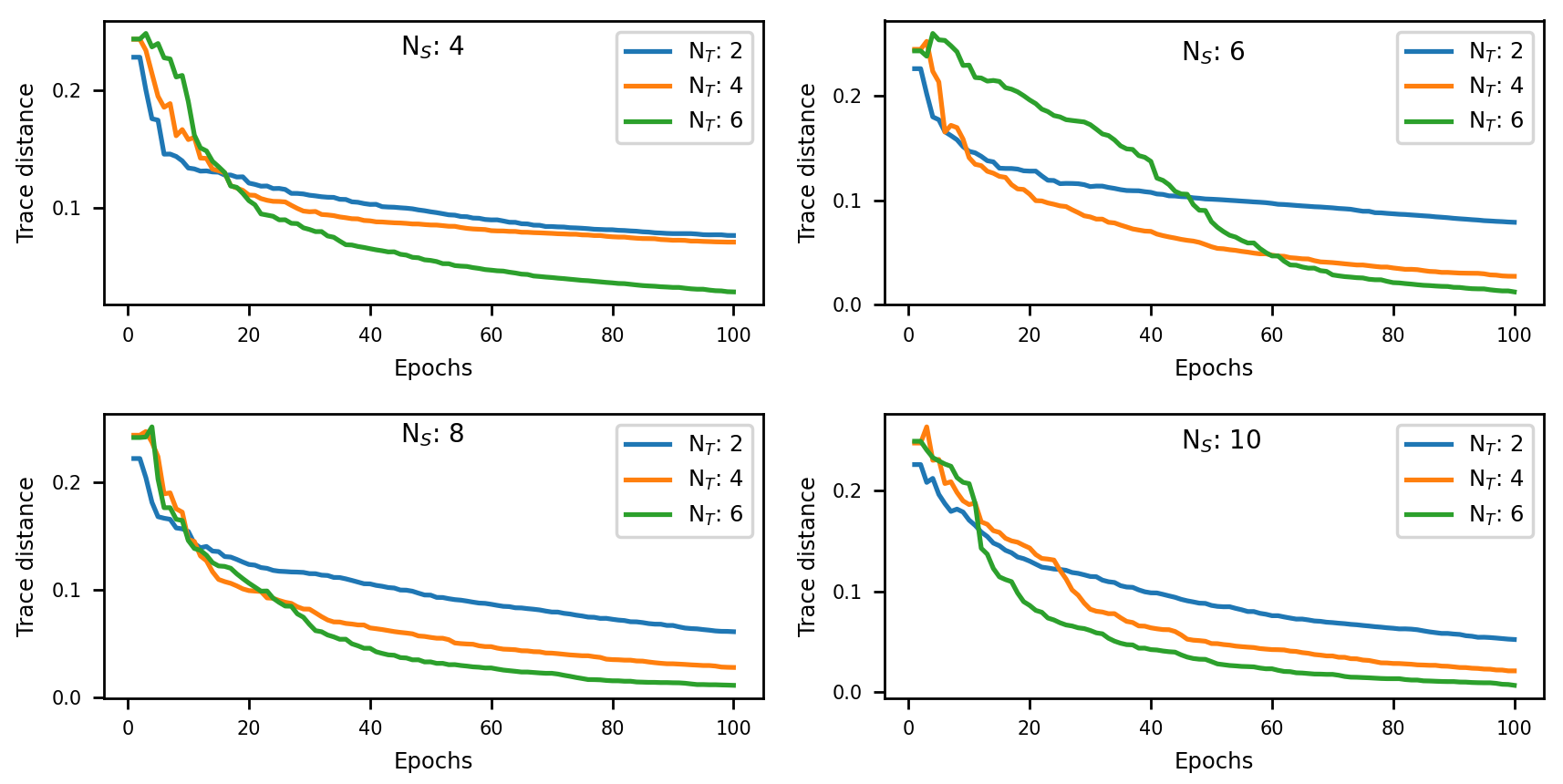} 
\caption{Variation of the trace distance between the true and the reconstructed Hamiltonian as a function of the number of epochs for fixed $N_O = 3$. It is evident that the Hamiltonian can be learned more efficiently if the expectation values of observables at higher time steps are also known for any number of starting quantum states. }
\label{fig_trace_same_nstates}
\end{figure} 
\begin{figure}[ht!]
  \centering 
\includegraphics[width=3.2in]{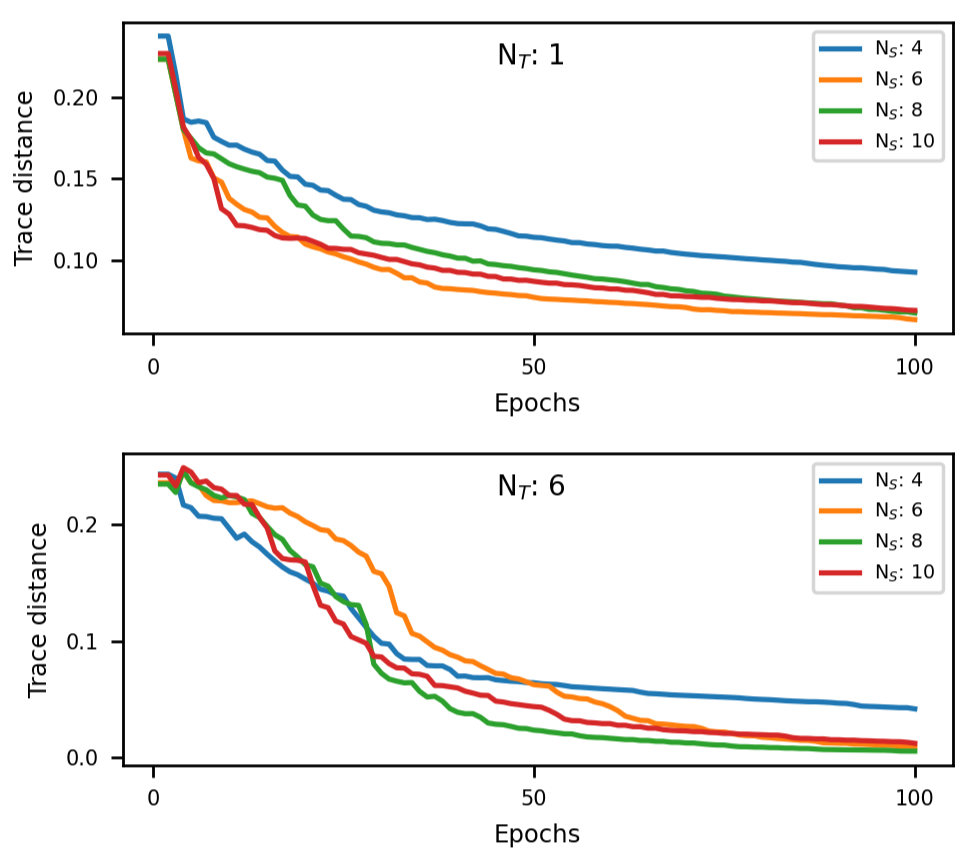} 
\caption{Convergence of the trace distance between the true and the reconstructed Hamiltonian is plotted to highlight the effect of increasing the number of starting quantum states when $N_O$ and $N_T$ are fixed.}
\label{fig_trace_same_tsteps}
\end{figure} 
Figure \ref{fig_trace_same_nstates} shows the variation of the trace distance with the number of epochs during the optimization process. Each subplot in Figure \ref{fig_trace_same_nstates} depicts the trace distance variation for a fixed number of starting quantum states but for a different number of time steps for which the observable's expectation values are known. For this plot we fixed the number of different observables of $\sigma_k^i\sigma_k^j$ to 3 i.e. we randomly selected expectation values of 3 $\sigma_z^i\sigma_z^j$ and 3 $\sigma_x^i\sigma_x^j$ correlation functions. As can be seen in the Figure, the convergence of the trace distance improves when observables corresponding to higher time steps are included in the optimization in all cases of $N_S$. To analyze the effect of the variation of $N_S$ on the accuracy of the reconstruction of the Hamiltonian, we show in Figure \ref{fig_trace_same_tsteps} the trace distance convergence for fixed $N_O = 3$, and $N_S$ being varied in each subplot for one particular $N_T$. We see that for $N_S \geq 6$ the converged trace distance is very close to 0 and therefore, although increasing the number of random quantum states does improve the performance up to a certain value of $N_S$ for all $N_T$ but beyond that, the accuracy is independent of increasing the quantum states.  \newline 
\begin{figure}[ht!]
  \centering 
\includegraphics[width=5.5in]{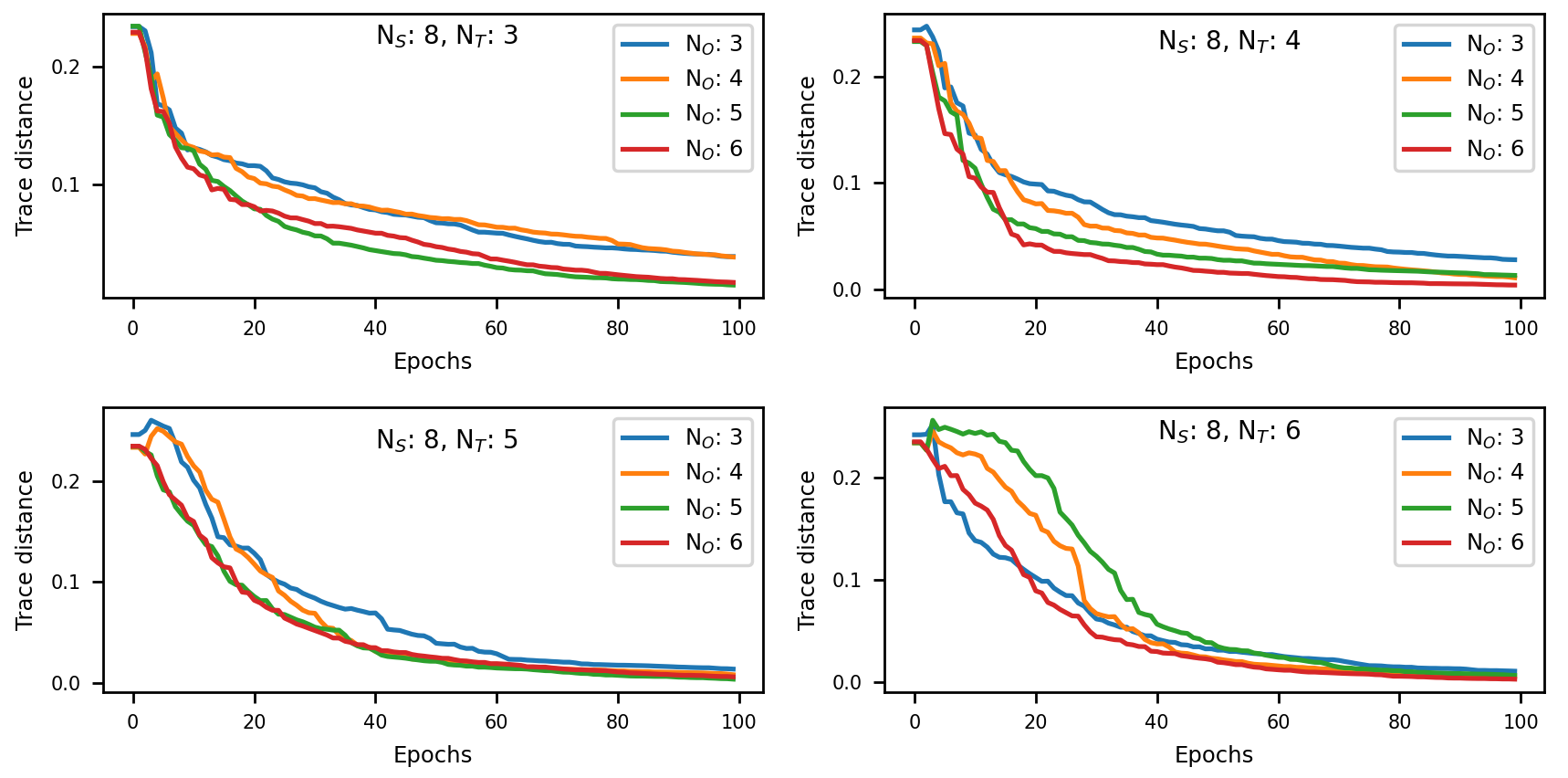} 
\caption{Plots of the trace distance upon varying the number of observables ($N_O$) considered for Hamiltonian learning. The convergence of the learned Hamiltonian to the true Hamiltonian improves upon increasing $N_O$ in all cases. It is also interesting to note that we can obtain a highly accurate Hamiltonian by using fewer observables for higher $N_T$ as is evident from the bottom plots.   }
\label{fig_trace_ops_vary}
\end{figure} 
The next step is to analyze the performance of our approach upon varying the number of observables (correlation functions) whose expectation values are considered in the Hamiltonian learning protocol. In Figure \ref{fig_trace_ops_vary}, we plot the trajectories of trace distance for the various number of observables ($N_O$) considered for a fixed $N_T$ and $N_S$ in each subplot. We see that the reconstructed Hamiltonian's trace distance with the true Hamiltonian converges to 0 faster upon increasing $N_O$ in the protocol. We also change $N_T$ in the different subplots to emphasize the advantage of using the time dynamics of observables for optimization. As can be seen in Figure \ref{fig_trace_ops_vary} that $T(H,K)$ approaches 0 upon optimization even for fewer observables upon including their expectation values for higher time steps. \newline
\begin{figure}[ht!]
  \centering 
\includegraphics[width=3.2in]{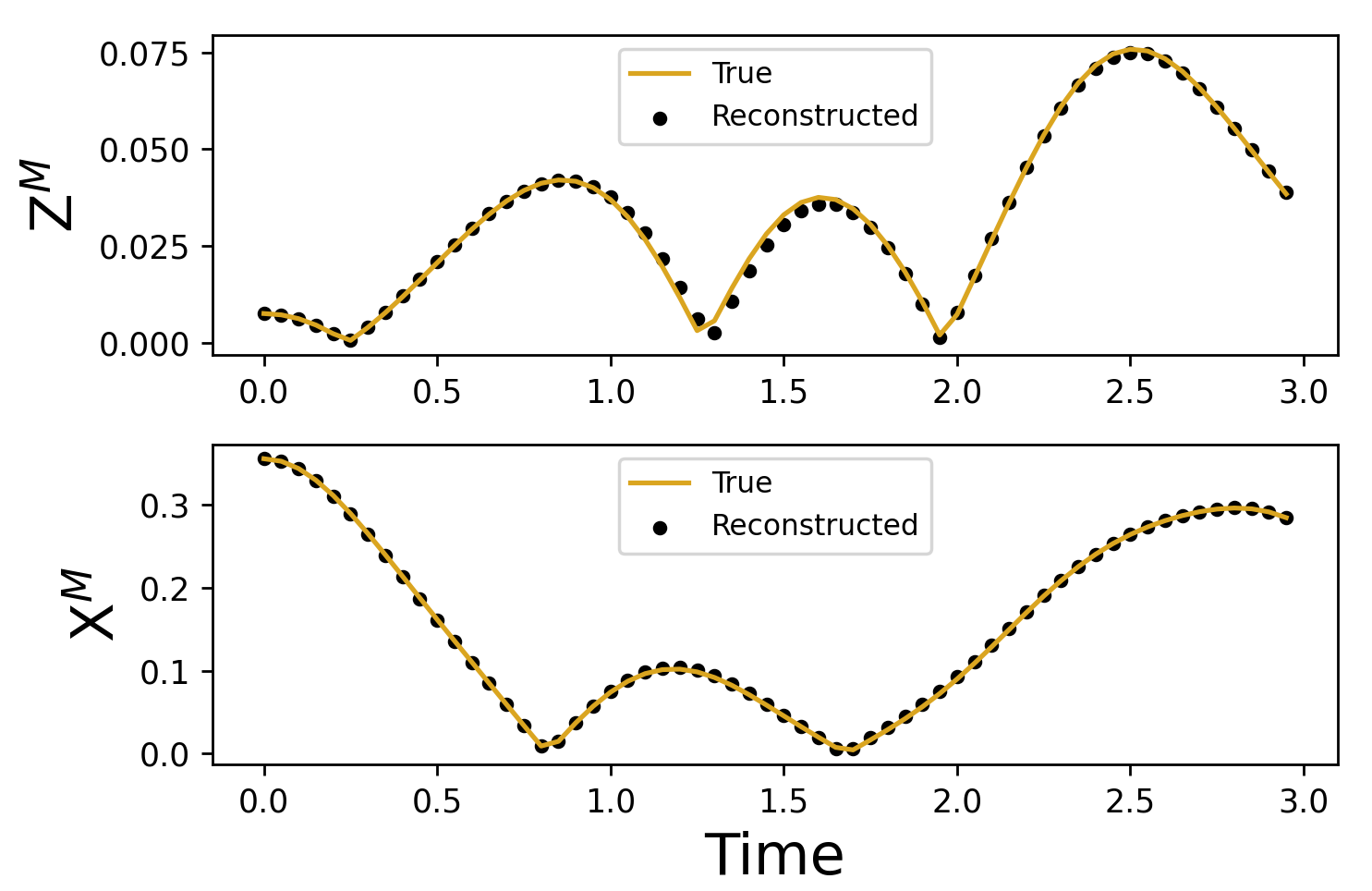} 
\caption{Time dynamics for $X^M$ and $Z^M$-magnetic field observables using the true and the reconstructed Hamiltonian for a general 2-local Hamiltonian with random coupling parameters.}
\label{fig_trace_observables}
\end{figure} 
Once the optimized set of parameters is obtained upon convergence, one of the ways to test the accuracy of learning is to evolve a quantum state under the true and the learned Hamiltonian separately and compare the dynamics of an observable under such an evolution. For our case, we considered the X and Z-magnetic field operators, denoted by $X^M$ and $Z^M$ respectively, and defined by $\hat{K} = \sum_p \sigma_k^p\sigma_i\sigma_i\sigma_i\sigma_i$ (\textit{k = x, z}). We computed the time dynamics of these operators using the evolution of a random quantum state under the true and the learned Hamiltonian. High overlap in the trajectories in Figure \ref{fig_trace_observables} validates that the proposed protocol is successful in the efficient reconstruction of an unknown Hamiltonian.

\subsection{Reconstructing Transverse Field Ising Model Hamiltonian}
The aforementioned results correspond to the learning of a general 2-local Hamiltonian with random coupling parameters. Since every quantum system has its own Hamiltonian so we need an empirical recipe for determining the optimized set of $N_O, N_S, N_T$. We show the reconstruction of the Transverse Field Ising Model (TFIM) Hamiltonian for the following two systems:

\subsubsection{Inhomogeneous 5-qubit Transverse Field Ising Model Hamiltonian} \label{inhomo}
\begin{figure}[ht!]
  \centering 
\includegraphics[width=5.5in]{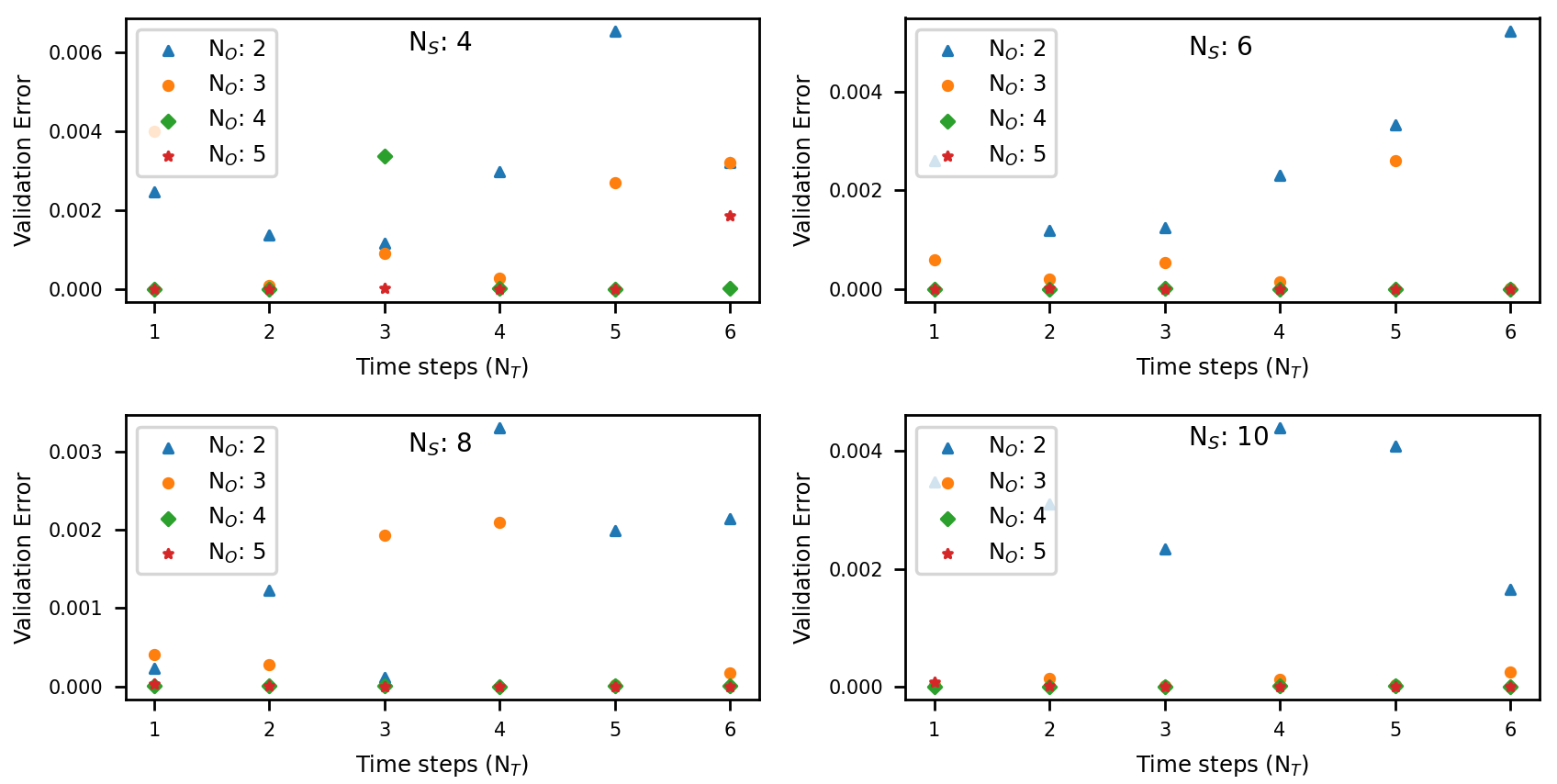} 
\caption{Validation error corresponding to the time dynamics of $Z^M$-magnetic field operator following the reconstruction of a 5-qubit TFIM Hamiltonian using the various values for $N_O, N_S, N_T$.}
\label{fig_valid_TFIM_5}
\end{figure}

Here we present a practical approach for Hamiltonian learning for the specific case of an inhomogeneous 5-qubit TFIM Hamiltonian, defined by $H=\sum_{i}c_i\sigma_x^i + \sum_{i,j=i+1}d_{ij}\sigma_z^i\sigma_z^j$, but is generalizable to any Hamiltonian. We inferred from the previous section that the time dynamics of various observables is an essential element for assisting the convergence of the trace distance. Therefore, from the perspective of experiments when the true Hamiltonian is unknown, one can start with obtaining mean values of a few observables using a few random starting quantum states but for higher time steps. Then our proposed optimization method can be applied to obtain the coupling parameters in the Hamiltonian. Since for practical implementation, we won't have access to the true Hamiltonian and so, one cannot calculate the trace distance to validate that the reconstructed Hamiltonian is close to the true Hamiltonian. However, what we can have access to is the time dynamics of another observable on the real system that is not used during the optimization process. We, therefore, define the concept of validation error as the mean square error between the time series expectation values of the observable calculated using a random quantum state evolved under the true Hamiltonian and the reconstructed Hamiltonian. Figure \ref{fig_valid_TFIM_5} shows the validation error corresponding to the $Z^M$-magnetic field operator for various values of $N_O, N_S, N_T$ for the considered TFIM Hamiltonian. Thus, the optimized set for $N_O, N_S, N_T$ can be obtained when the validation error is below a certain threshold as per the requirement of the experiment.  

\begin{figure}[ht!]
  \centering 
\includegraphics[width=3.2in]{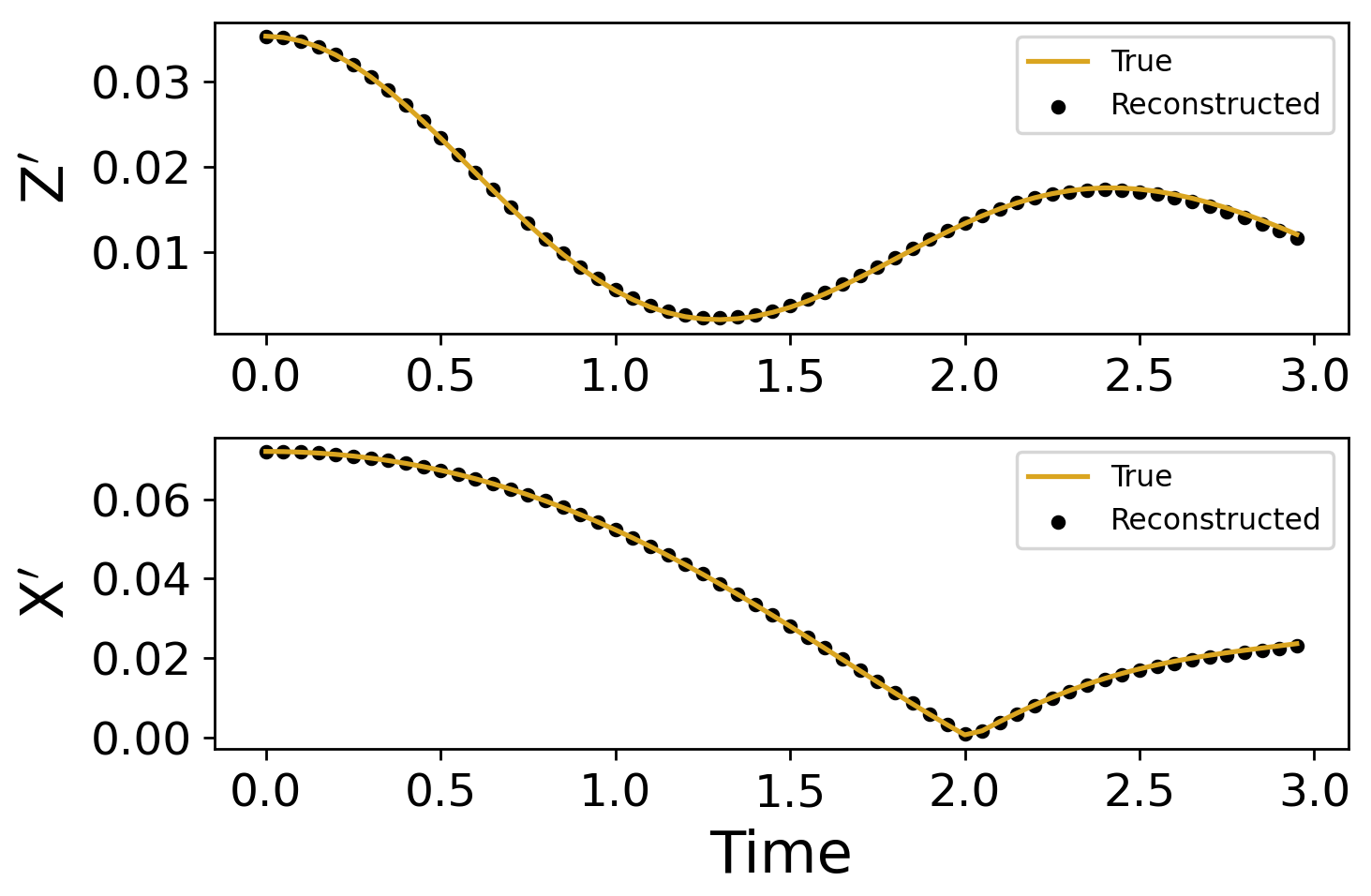} 
\caption{Time dynamics of 3-point correlation functions $X^\prime$ and $Z^\prime$ for 5-qubit TFIM Hamiltonian. The learned Hamiltonian is obtained using $N_O = 3, N_S = 8, N_T = 5$ for which the validation error is below 1e-6.}
\label{fig_obs_TFIM_5}
\end{figure} 
For the inhomogeneous 5-qubit TFIM we obtain the converged coupling parameters corresponding to $N_O, N_S, N_T$ for which the validation error is below 1e-6. The true and the learned Hamiltonian is then used to evolve a random quantum state to obtain the time series expectation values of the observables $X^\prime$ and $Z^\prime$ which are the 3-point correlation functions $\sigma_x^i\sigma_x^j\sigma_x^k$ and $\sigma_z^i\sigma_z^j\sigma_z^k$. The performance efficacy of our method can be validated from Figure \ref{fig_obs_TFIM_5} which shows the overlap at each time step in the dynamics of these observables using the true and the reconstructed Hamiltonian.   

\subsubsection{Homogeneous 10-qubit Transverse Field Ising Model Hamiltonian}
\begin{figure}[ht!]
  \centering 
\includegraphics[width=3.2in]{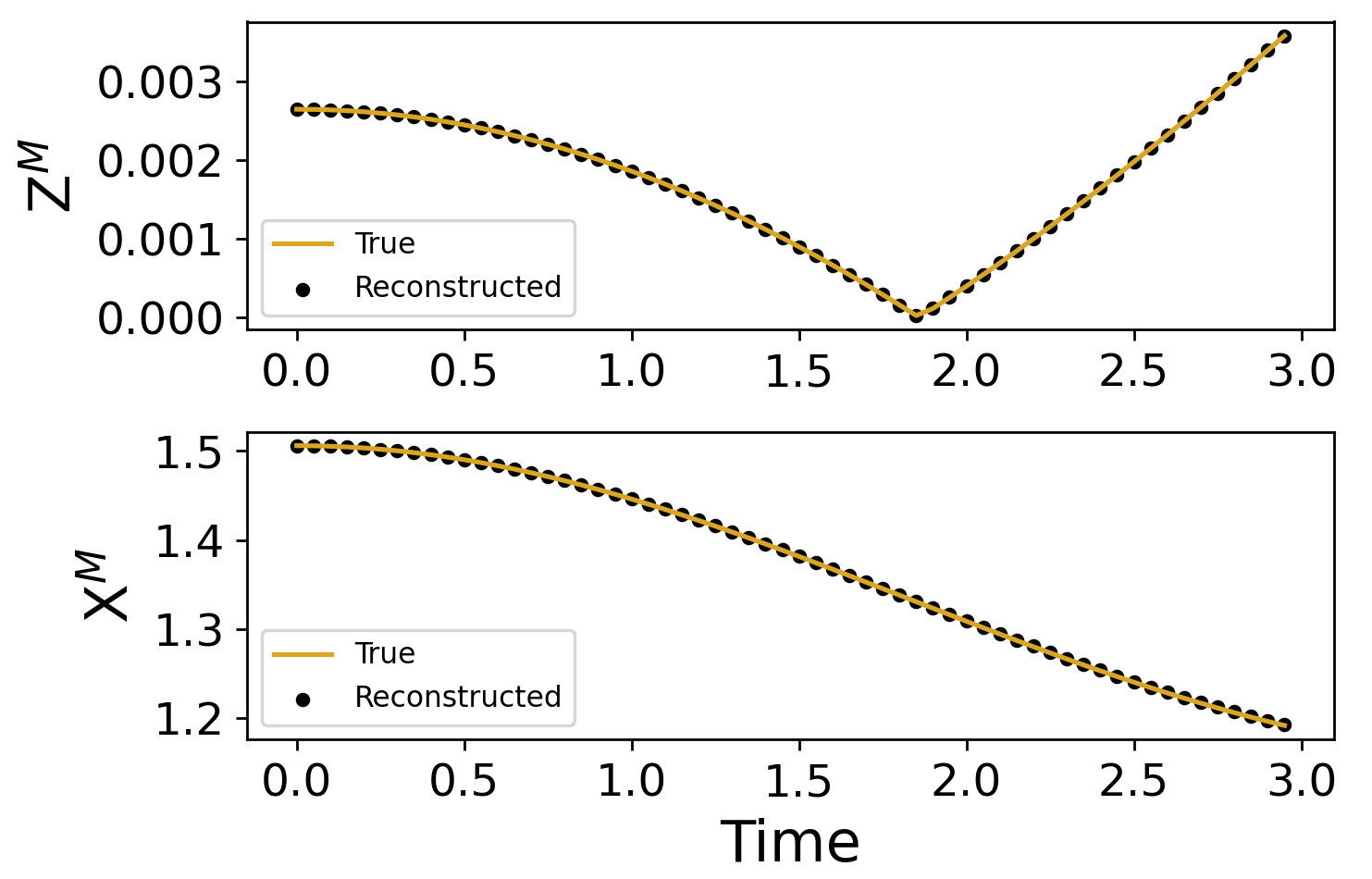} 
\caption{Time dynamics of $X^M$ and $Z^M$-magnetic field observables for 10-qubit homogeneous TFIM Hamiltonian. The learned Hamiltonian is obtained using $N_O = 2, N_S = 2, N_T = 1$ for which the validation error is below 1e-6.}
\label{fig_obs_10_qubits}
\end{figure}

The quantum Ising chain is a classic domain for testing various ideas and methodologies from statistical mechanics and consequently, forms a classic candidate to experimentally test the efficiency of any hamiltonian learning formalism. For the learning of a homogeneous TFIM hamiltonian, defined by $H= h\sum_{i}\sigma_x^i + J\sum_{i,j=i+1}\sigma_z^i\sigma_z^j$, we considered a quantum system comprising of 10-qubits. Since there are only 2 parameters (h, J) that need to be learned, the parameterized hamiltonian from the proposed protocol converges quickly to the true Hamiltonian. We again chronologically vary the set of ($N_T, N_S, N_O$) during optimization to procure that set for which the validation error is below 1e-6 following which the training parameters are obtained and used for predicting the dynamics of certain unknown observables. Figure \ref{fig_obs_10_qubits} shows the dynamics of X and Z-magnetic fields' observables under the true and the learned Hamiltonian for various time steps which again validates the learning approach.    

\subsection{Quantum State Learning}
\begin{figure}[ht!]
  \centering 
\includegraphics[width=5.5in]{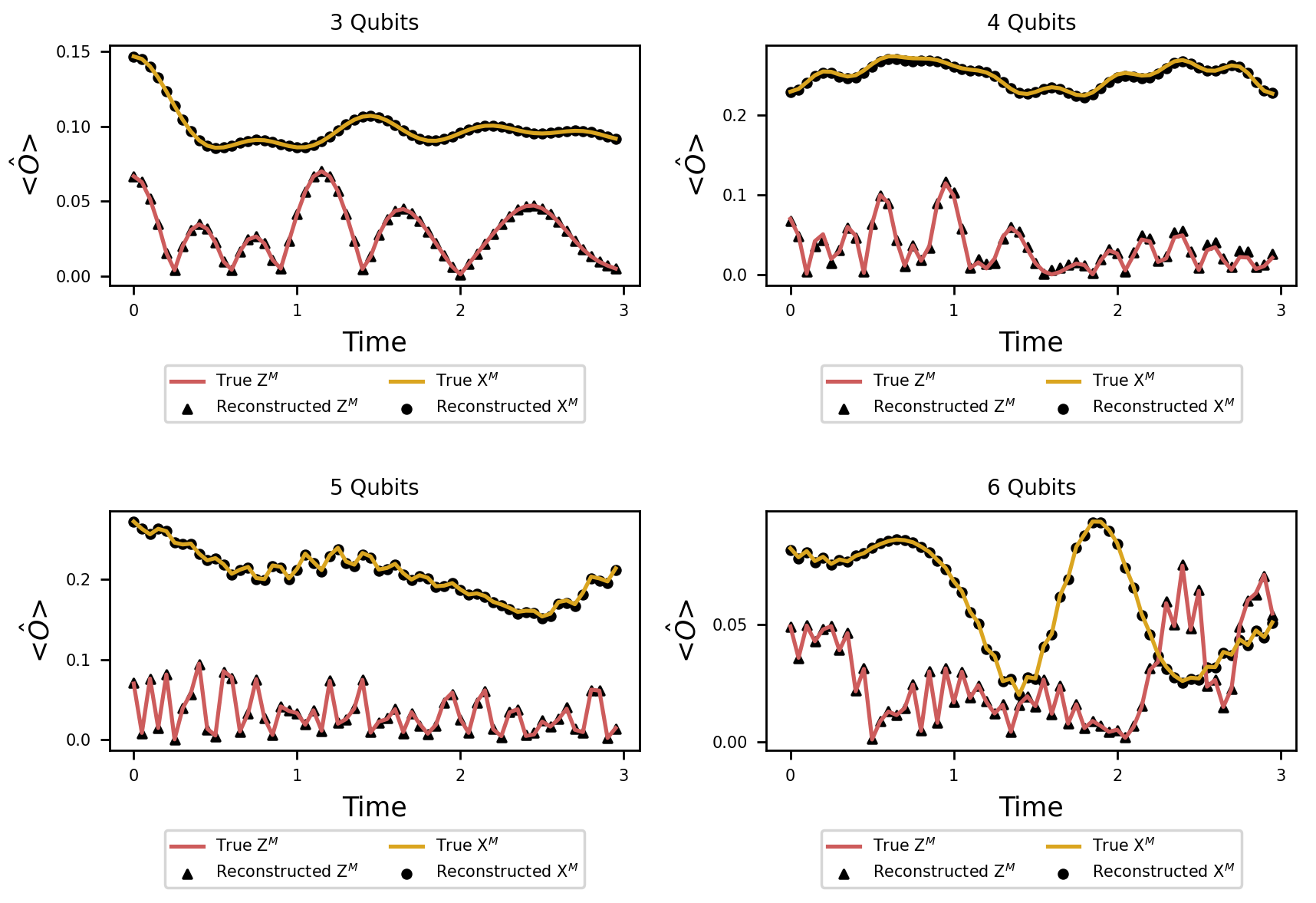} 
\caption{Time dynamics of $X^M$ and $Z^M$-magnetic field observables under the evolution of a random Hamiltonian using the true and the learned quantum states for various \textit{n}-qubit quantum states.}
\label{fig_state_valid}
\end{figure}

Efficient characterization of a pure quantum state by learning the parameters of a parameterized quantum circuit has recently been gaining much popularity within the domain of quantum state learning owing to their use of shallow quantum circuits with low circuit depth. Here, we extend our idea of Hamiltonian learning and propose to learn the quantum state of a system using PQC based on the knowledge of the time dynamics of various observables obtained from the evolution of the state under different random Hamiltonians. We start with a hardware-efficient ansatz whose parameters are variationally optimized to satisfy the constraints of the problem. We show the learning of random \textit{n}-qubit quantum states for up to 6 qubits. \newline
\begin{figure}[ht!]
  \centering 
\includegraphics[width=5.5in]{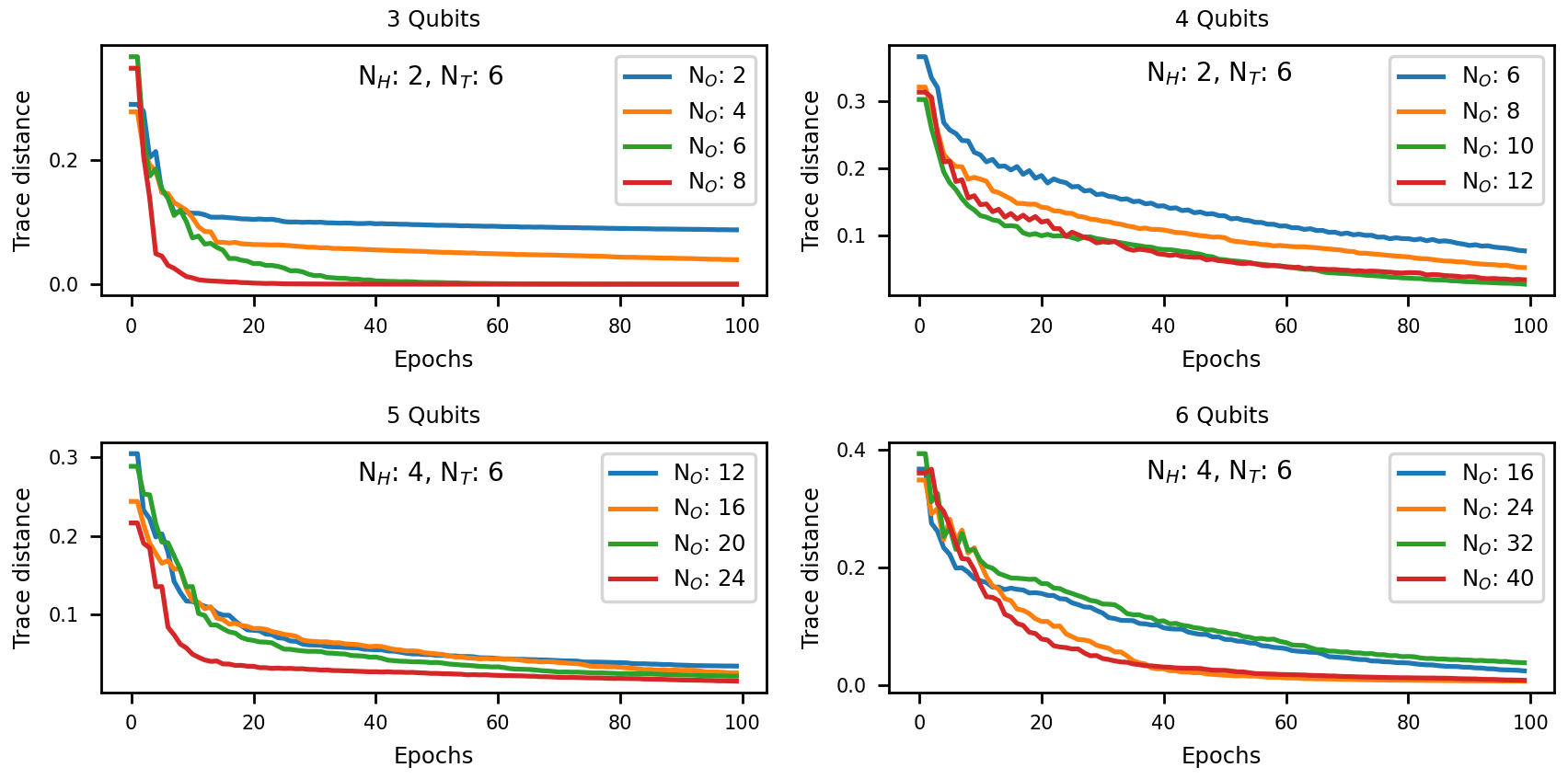} 
\caption{Trace distance between the true and the learned quantum state for various \textit{n}-qubit quantum states using the time dynamics approach for fixed $N_H$, $N_T$, and varying $N_O$. Even though an informationally incomplete set of observables are used but still the time dynamics approach is able to reconstruct the quantum state close to the true state, thereby validating the proposed formalism. }
\label{fig_state_qubits}
\end{figure} 
As opposed to varying the number of random initial quantum states, for state learning, we start with a parameterized quantum state and evolve it under different Hamiltonians chosen at random for various time steps and compute the expectation values of different observables. The evolution operator for the Hamiltonian can be implemented on a quantum circuit using Trotterization as discussed in Appendix \ref{trotter}. However, for the proof of concept, we numerically evaluate the evolution operator since we are dealing with only a few qubits. Here, the choice of observables whose time dynamics we consider in the optimization protocol consists of the operators corresponding to the probabilities and coherences of the density matrix. Ideally, the number of observables required for an accurate and unique representation of a quantum state grows exponentially with the system size. Also, not all observables can be obtained with high fidelity in an experiment. This is why our method can prove to be advantageous as if we can obtain the time dynamics of fewer easily measurable observables then our protocol can learn the quantum state of a system with high fidelity. The cost function, in this case, is similar to that in Hamiltonian learning and is discussed in Section \ref{method}. We minimize the mean square error between the time dynamics of the observables obtained from the true state and the reconstructed state. The comprehensive results corresponding to the variation of the trace distance (T($\rho,\sigma$) = $\frac{1}{2}||\rho-\sigma||_1$) between the true and the learned state during optimization is shown in Appendix \ref{state}. We vary the number of random Hamiltonians ($N_H$) under which the quantum state is evolved, the number of time steps ($N_T$) for evolution, and the number of observables ($N_O$) to obtain the converge set of parameters for which the trace distance  is within an error threshold. However as before, in an experiment, we won't have access to the trace distance as the true state is unknown so we can again go back to the concept of a validation error. As we learn the quantum state upon optimization for one set of ($N_T, N_H, N_O$) we then evolve it under a known Hamiltonian and obtain the time dynamics of a new observable not used during learning and compute the validation error. Depending on the bound of the error threshold, one can choose the parameters for ($N_T, N_H, N_O$). In Figure \ref{fig_state_qubits} we see the variation of the trace distance with the number of epochs for quantum states prepared on various \textit{n}-qubit quantum systems. To show the accuracy of reconstructing the quantum state, we compute the evolution of the reconstructed state and the true state under a random Hamiltonian and plot the time dynamics of X and Z-magnetic fields' observables in Figure \ref{fig_state_valid}.

\subsection{Generalized Hamiltonian learning for SU(3) group}
Quantum information processing of qutrit systems, consisting of three-dimensional quantum states, is of special significance and has drawn considerable attention both theoretically and experimentally \cite{li2013geometry}. For example, the nitrogen-vacancy center in diamond \cite{gardill2020fast}, three ground states of trapped \ce{^{171}Yb+} ion, etc \cite{lucarelli2002chow}. can serve as a qutrit system on which various quantum operations can be performed. The geometry of the special unitary  SU(3$^n$) group can be used to define a quantum gate on \textit{n}-qutrit quantum state. The use of Lie algebraic operations provides a technical advantage as certain classes of molecules can be described by common Hamiltonian, differing only by linear parameters \cite{iachello1995algebraic}. In Appendix \ref{gellmann} we derive the mathematical expression for the calculation of analytical gradients for SU(3) group Hamiltonians defined as $H = \sum_{i=1}^8 c_i\lambda_i$, where $\lambda_i$ correspond to Gell-Mann matrices that span the Lie algebra of SU(3) group. Based on the cost function defined in Eq. (\ref{cost1}), the gradients with respect to the coefficients $\{c_p\}$ are: 
\begin{equation}
    \frac{\partial Cost}{\partial c_p} = 2 \sum_{\alpha ,i,k}^{N_O, N_S, N_T}\bra{\psi_i}\frac{\partial U^\dagger(k\Delta t) O_\alpha U(k\Delta t)}{\partial c_p}\ket{\psi_i} \label{Full_grad}
\end{equation}
Using the analytical gradients as shown in Appendix \ref{gellmann}, it must be emphasized that the protocol so engendered is cost-effective as for a given choice of $(O_\alpha, \psi_i)$ (corresponding to each term within Eq. (\ref{Full_grad})) the set of quantum measurements necessary to construct Eq. (\ref{Full_grad}) $\forall k$ time steps can be expressed in terms of 
9 measurements for $SU(3)$ (for a general $SU(N)$ this number will be $N^2$)  evaluated from the quantum circuit using the initial state $|\psi_i\rangle$ alone provided $O_\alpha \in \{\lambda_i\}_{i=1}^{9}$. For each of these 9 measurements, the primitive observables that need to be measured are the generators $\{\lambda_i\}_{i=1}^{9}$ defined before. One must note that this obviates the need to perform repetitive measurements using the time-evolved version of $U(k\Delta t)$ for all of the successive $N_T$ time steps as is usually necessitated for systems wherein such a Lie-algebraic structure is not present. In other words, the total number of measurements to construct Eq. (\ref{Full_grad_SU3}) in its entirety will then simply be $9N_ON_S$ which is independent of $N_T$. Thus, using the structural properties of the SU(\textit{N}) group, we can further extend this approach efficiently for Hamiltonian learning of multi-dimensional qudit quantum systems as well. 

\section{Concluding remarks}
Measuring the time dependence of the mean values of observables enables us to solve two complementary problems. First and foremost we seek to reconstruct the Hamiltonian that generated the dynamics of various quantum systems. On the flip side we characterize the state of the evolving system. We provide numerical results showing how the method works. We used Trotterization to implement the evolution operator onto the quantum circuit and the entire workhorse of our proposed formalism is centered around variational optimization using shallow-depth quantum circuits and thus, is easily implementable on a near-term quantum device. We presented a comprehensive list of results corresponding to the Hamiltonian learning of various 2-local Hamiltonians as well as state learning of quantum systems comprising up to 6 qubits where we illustrate the significance of using time dynamics within the optimization process. The simple nature of our proposed approach can find its application in a variety of different areas where the dynamics of various observables need to be studied. As a subsidiary validation we examine a special (but general) case where the problem can be solved analytically. In Appendix \ref{gellmann}, we propose to learn the Hamiltonian of multi-dimensional qudit systems by exploiting the structural properties of SU(\textit{N}) groups and specifically show the analytical gradient calculation procedure for qutrit systems defined by SU(3) geometry. We also intend to further investigate the utility of the methods developed here for studying time series datasets in typical machine learning algorithms that are commonly tackled using long short-term memory models (LSTM) \cite{LSTM} and recurrent neural network models (RNN) \cite{Sherstinsky_2020}. 
% \begin{itemize}
%     \item summarise the findings
%     \item what future work can be done here
%     \item where might these findings be helpful
%     \item give treat for teammates :)
% \end{itemize}    

\section{Appendix}
\label{sec:Appendix}
\subsection{Trotterized time evolution of a Hamiltonian} \label{trotter}
We consider the Heisenberg model on a 1D lattice with nearest-neighbor interactions whose Hamiltonian can be defined as $H = H_{xx} + H_{yy} + H_{zz}$ where
\begin{equation*}
H_{xx} = \sum_{i} J \sigma_x^{i}\sigma_x^{i+1} \\
H_{yy} = \sum_{j} J \sigma_y^{j}\sigma_y^{j+1} \\
H_{zz} = \sum_{k} J \sigma_z^{k}\sigma_z^{k+1}        
\end{equation*}
We are interested in evaluating the evolution operator $U = exp(-iHt)$ but the non-commutation of the Pauli operators doesn't allow the operator \textit{U} to be written as the product of simpler exponentials as $U \neq exp(-iH_{xx}t)exp(-iH_{yy}t)exp(-iH_{zz}t)$. This is where Trotterization can be used to approximate \textit{U} by splitting it into the product of simpler exponentials using \textit{N} trotter steps giving an approximation up to second order in $t/N$ as $U \approx \big[ exp(-iH_{xx}t/N)exp(-iH_{yy}t/N)exp(-iH_{zz}t/N)\big]^{N}$. Here, we show Trotterization specifically for the nearest neighbor XYZ Hamiltonian but it can be generalized for others. \newline
\begin{figure}[ht!]
  \centering 
\includegraphics[width=4.2in]{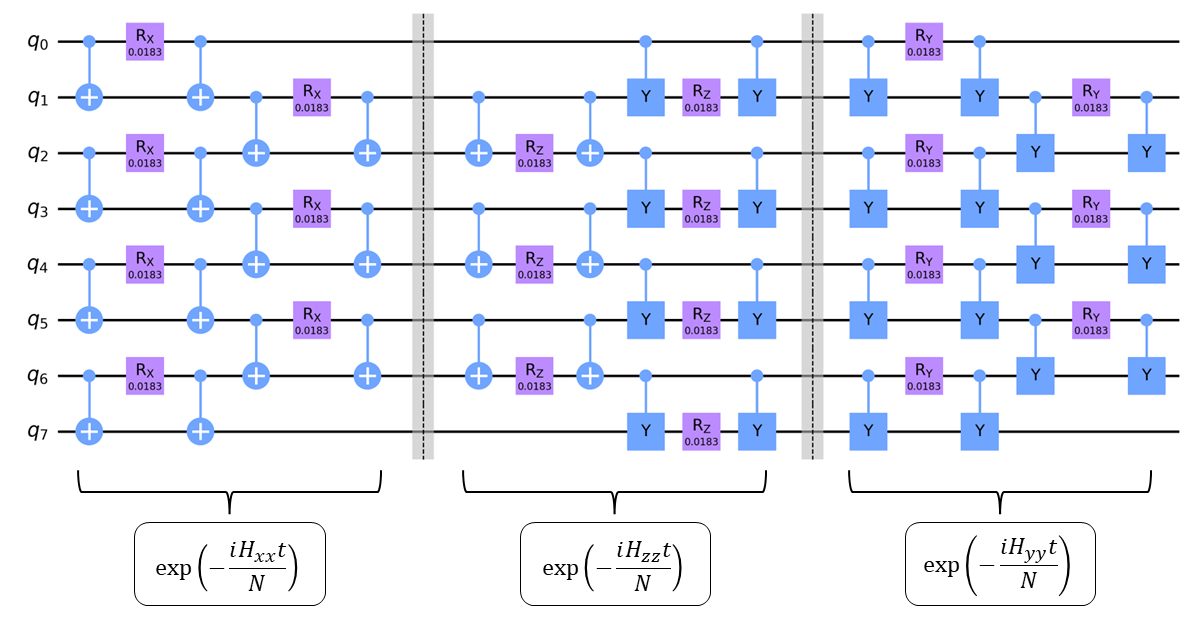} 
\caption{Quantum circuit corresponding to one step of Trotterization for the time evolution of a Hamiltonian comprising of only nearest-neighbors interactions.}
\label{fig_trotter_full}
\end{figure} 
\begin{figure}[ht!]
  \centering 
\includegraphics[width=3.6in]{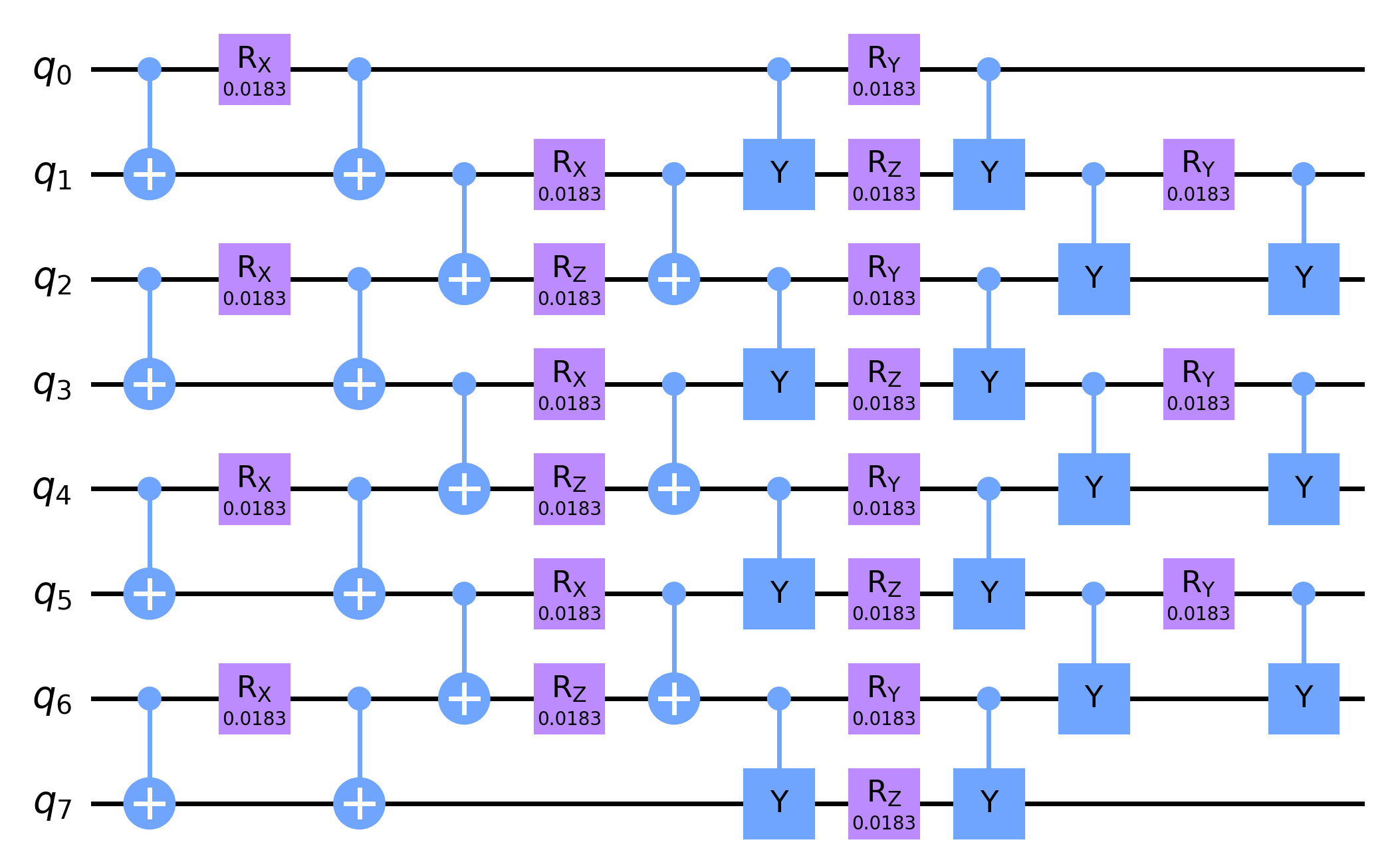} 
\caption{Quantum circuit for one step Trotterization obtained by eliminating the unnecessary CNOT and CY gates from Figure \ref{fig_trotter_full} to reduce the circuit depth from 18 to 12.}
\label{fig_trotter}
\end{figure} 
Let us consider a Hamiltonian as defined above for an 8-qubit quantum system. For nearest neighbor coupling each qubit couples with a maximum of 2 qubits. We can thus split the execution into 2 layers such that qubits in the first layer couple with exactly one other qubit. This ensures that the overall depth of the circuit is constant and independent of the number of qubits. It scales only with the degree of the vertex which is 2 (constant) here. Figure \ref{fig_trotter_full} shows the quantum circuit implementation of one-step Trotterization, using only 1 and 2-qubit quantum gates, that can be repeated \textit{N} times for full-time evolution. The number of 2 qubit gates scales as 4\textit{n}-4 for the construction as shown in Figure \ref{fig_trotter}. The advantage of using this scheme of Trotterization is evident from the circuit diagram in Figure \ref{fig_trotter_full} as essentially the CNOT and CY gates around the shown barrier are not required and so the circuit depth can be reduced as can be seen in Figure \ref{fig_trotter}. 

\subsection{Gradients calculation on quantum circuit} \label{gradient}

Since our ansatz makes use of Pauli rotation gates we instead can compute them analytically. The most general Pauli rotation operator is given by $exp(-i \vec{n}.\vec{\sigma}\theta)$, where $\vec{\sigma} = (I,\sigma_x ,\sigma_y, \sigma_z)$ ,$\vec{n} =(n_0,n_1,n_2,n_3)$ and $\theta$ is the angle of rotation. Using the identity that $exp^{-i \vec{n}.\vec{\sigma}\theta} = cos(\theta)n_0I + i sin(\theta)(n_1\sigma_x + n_1\sigma_y + n_1\sigma_z)$ one can straightforwardly derive an expression for the gradients,
\begin{equation}
    \begin{split}
        \frac{d\bra{\psi}H\ket{\psi}}{d\theta_i} & = \frac{d\bra{\psi}}{d\theta_i}H\ket{\psi} + \bra{\psi}H\frac{d\ket{\psi}}{d\theta_i} \\
        & = \mathrm{Re}(\frac{d\bra{\psi}}{d\theta_i}H\ket{\psi}) \quad \mathrm{(Since \; H \; is \; hermitian)}\\
    \end{split}
\end{equation}
Now since our unitaries are a function of $R_y(\theta)$ we have,
\begin{equation}
    \begin{split}
    \frac{d R_y}{d\theta} & = \frac{d e^{i \sigma_y \theta/2}} {d\theta} = \frac{1}{2} i \sigma_y  e^{i \sigma_y \theta/2} \\
    & = \frac{1}{2} e^{i \sigma_y \pi/2} e^{i \sigma_y \theta/2}  = \frac{1}{2} R_y(\pi) R_y(\theta) \\
    & = \frac{1}{2} R_y(\theta + \pi)  
    \end{split}
\end{equation}

Thus computing the gradients of the cost function with respect to parameters of Pauli rotation gates can be straightforwardly evaluated on the circuit with $\pi$ shift of the corresponding parameter. Alternatively for a unitary $U=exp(-i\frac{\theta}{2}P_i)$, where $P_i$ is a pauli operator, one could use the expression from parameter shift rules \cite{schuld2019evaluating} to compute the gradient of a cost function $C(\alpha,\beta) = \bra{\psi}U^{\dag}(\alpha,\beta)\;O\;U(\alpha,\beta)\ket{\psi}$ given by,

\begin{equation}
    \frac{dC(\alpha,\beta)}{d\alpha} = \frac{1}{2} \big[  C(\alpha + \frac{\pi}{2},\beta) - C(\alpha - \frac{\pi}{2},\beta) \big]
\end{equation}

% Alternatively, if the observables being measured are expressed in a complete basis of the Generators that span the Unitary group of the same dimension, one can use the commutators to simplify it at the level unitaries, without having to think about the circuit decomposition. In this case, the derivative of the expectation value is given by,

% \begin{equation}
% \centering
%     \frac{d\bra{0}U^{\dag}(\vec{c})\Tilde{G}U(\vec{c})\ket{0}}{dc_k} = \bra{0}U(\vec{c})i[G_k,\Tilde{G}]U(\vec{c})\ket{0}
% \end{equation}

% where $U(\vec{c}) = exp(-i\sum c_kG_k)$. Since the generators of SU(N) form a closed group, one can express $[G_a,G_b] =i \sum_c f_{abc}G_c$ using the structure coefficients. Hence one can directly compute the expectation value of the observables $G_c$, to get all the necessary gradients with respect to any of the coefficients that multiply the generators. This method however relies on the fact that there is an efficient mapping of $G_i$'s to the computational gate set being used (in the qubit case, Pauli rotation gates). 

\subsection{Cost function and convergence criteria for Hamiltonian learning}
\begin{figure}[ht!]
  \centering 
\includegraphics[width=6.2in]{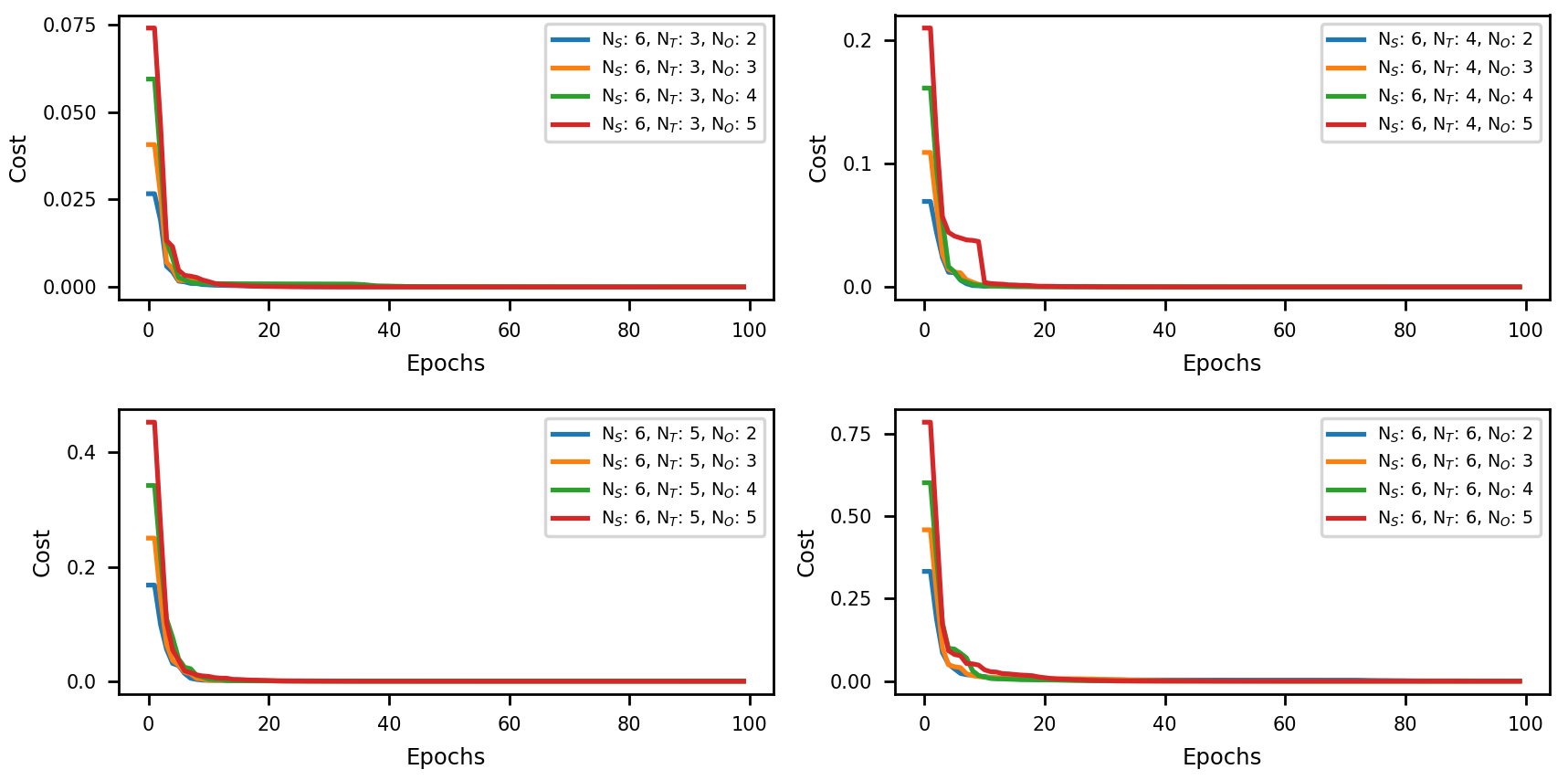} 
\caption{The mean square error (cost) as a function of the epochs for various sets of ($N_S$, $N_T$, $N_O$).}
\label{fig_trace_cost}
\end{figure}

\begin{figure}[ht!]
  \centering 
\includegraphics[width=4.2in]{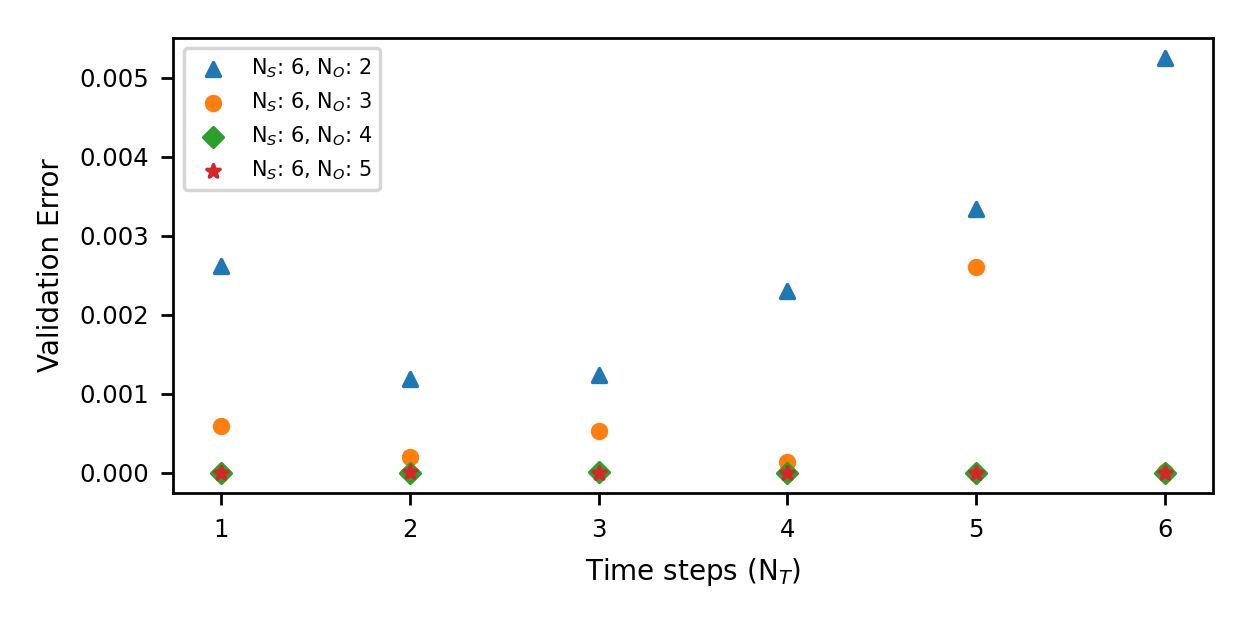} 
\caption{The validation error plot showing that even though the cost function converges to 0 in Figure \ref{fig_trace_cost} for all sets of ($N_S$, $N_T$, $N_O$) yet the corresponding learned Hamiltonian is close to the true Hamiltonian for those sets for which the validation error converges to 0.}
\label{fig_valid_error_app}
\end{figure}

The recipe for Hamiltonian learning in our case involves the experimental knowledge of the time dynamics of expectation values of correlation functions starting with multiple random quantum states. The number of correlation functions and random initial states required for Hamiltonian reconstruction depends on the error threshold $\epsilon$, that the experimentalist is interested in, between the true and the reconstructed Hamiltonian and can be evaluated using the validation error. As discussed in Section \ref{method}, let $O_{\alpha}(k \Delta t)$ refer to the measurement of observable $O_{\alpha}$ made at timestep $k$, where $\Delta t$ refers to the time interval of each timestep. Then, the cost function is defined as the 2-norm function between $O_{\alpha}(k \Delta t)$ and $O^{obs}_{\alpha,i}(k \Delta t)$ over the states $\ket{\psi_i}$:   
\begin{equation}
    \mathrm{Cost} = \sum_{\alpha ,i,k} (\bra{\psi_i}U^{\dagger}(k\Delta t) \; O_{\alpha} \; U(k\Delta t)\ket{\psi_i} - O^{obs}_{\alpha,i,k}) ^2 \label{cost_func}
\end{equation} 
where $U(k\Delta t)$ is the unitary operator that evolves the quantum state under a Hamiltonian \textit{H}. As shown in Section \ref{result}, the variables in our proposed protocol that affect the accuracy of Hamiltonian learning are the number of quantum states ($N_S$), the number of time steps ($N_T$) up to which we consider the time dynamics, and the number of observables ($N_O$) whose time dynamics we consider in the cost function. In section \ref{result} we have shown the results corresponding to the trace distance between the true and the learned Hamiltonian and also defined the validation error for choosing the correct set of ($N_S$, $N_T$, $N_O$). In Figure \ref{fig_trace_cost} the cost function is plotted as a function of the epochs for different sets of ($N_S$, $N_T$, $N_O$), and the corresponding validation error is plotted in Figure \ref{fig_valid_error_app} for the inhomogeneous 5-qubit TFIM Hamiltonian defined in Section \ref{inhomo}. As can be seen the mean square error loss i.e. the cost converges to 0 in all cases yet the validation error is significantly high for accurate learning in various cases. This is because even though the cost converges to 0 for fewer starting states, observables, and time steps but that information is insufficient to have accurate learning of the Hamiltonian. Thus the concept of validation error is important as based on it we can check whether or not the learned Hamiltonian, obtained from a given set of ($N_S$, $N_T$, $N_O$), is close to the true Hamiltonian.  

\subsection{Quantum state learning from time dynamics of observables} \label{state}
\begin{figure}[ht!]
  \centering 
\includegraphics[width=6.2in]{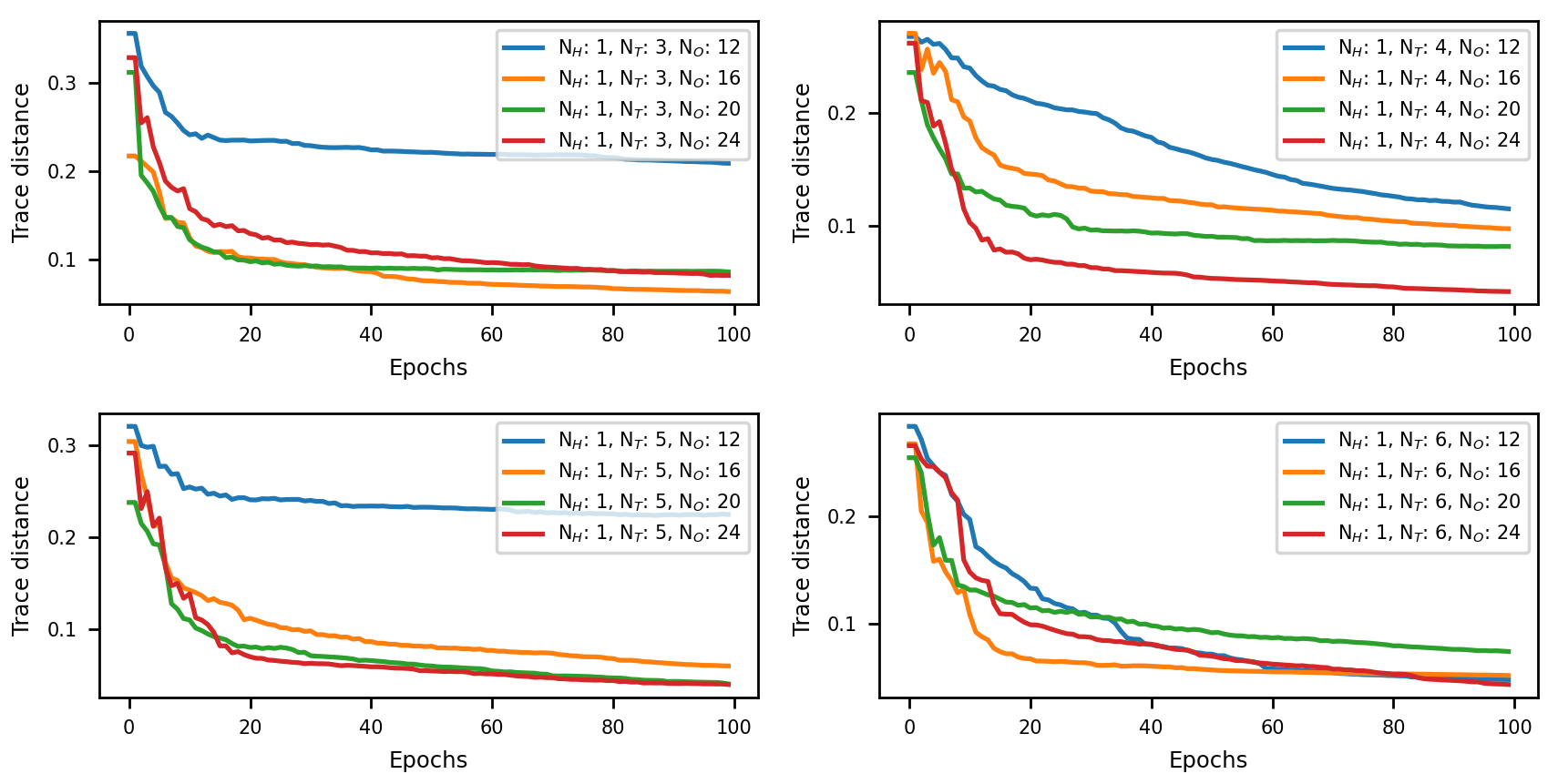} 
\caption{Varying the number of observables included during optimization shows that the accuracy of state learning improves upon increasing $N_O$. However, we can achieve similar accuracy using fewer observables if we have the knowledge of observables at higher time steps.}
\label{fig_fstrace_ops_vary}
\end{figure}

A natural extension of the idea of using time dynamics of various observables for Hamiltonian learning can be applied to solving the inverse problem of learning an unknown quantum state of a given system. Here, instead of starting with multiple random quantum states, we have at our disposal multiple random Hamiltonians ($N_H$) that are used for the evolution of the quantum state and the expectation values of different observables ($N_O$) are recorded at various time steps ($N_T$). The observables here correspond to the probabilities and coherences of the \textit{n}-qubit quantum system: 
\begin{eqnarray}
    &\hspace*{0.001cm}&\{\ket{1}\bra{1},\ket{2}\bra{2},\ket{3}\bra{3},\ket{4}\bra{4},(\ket{1}\bra{2}\pm\ket{2}\bra{1}),(\ket{1}\bra{3}\pm\ket{3}\bra{1}),(\ket{1}\bra{4}\pm\ket{4}\bra{1}), \nonumber \\
    &\hspace*{0.001cm}&(\ket{2}\bra{3}\pm\ket{3}\bra{2}), (\ket{2}\bra{4}\pm\ket{4}\bra{2}), (\ket{3}\bra{4}\pm\ket{4}\bra{3}) \} \label{basis1}
\end{eqnarray}
\begin{figure}[ht!]
  \centering 
\includegraphics[width=6.2in]{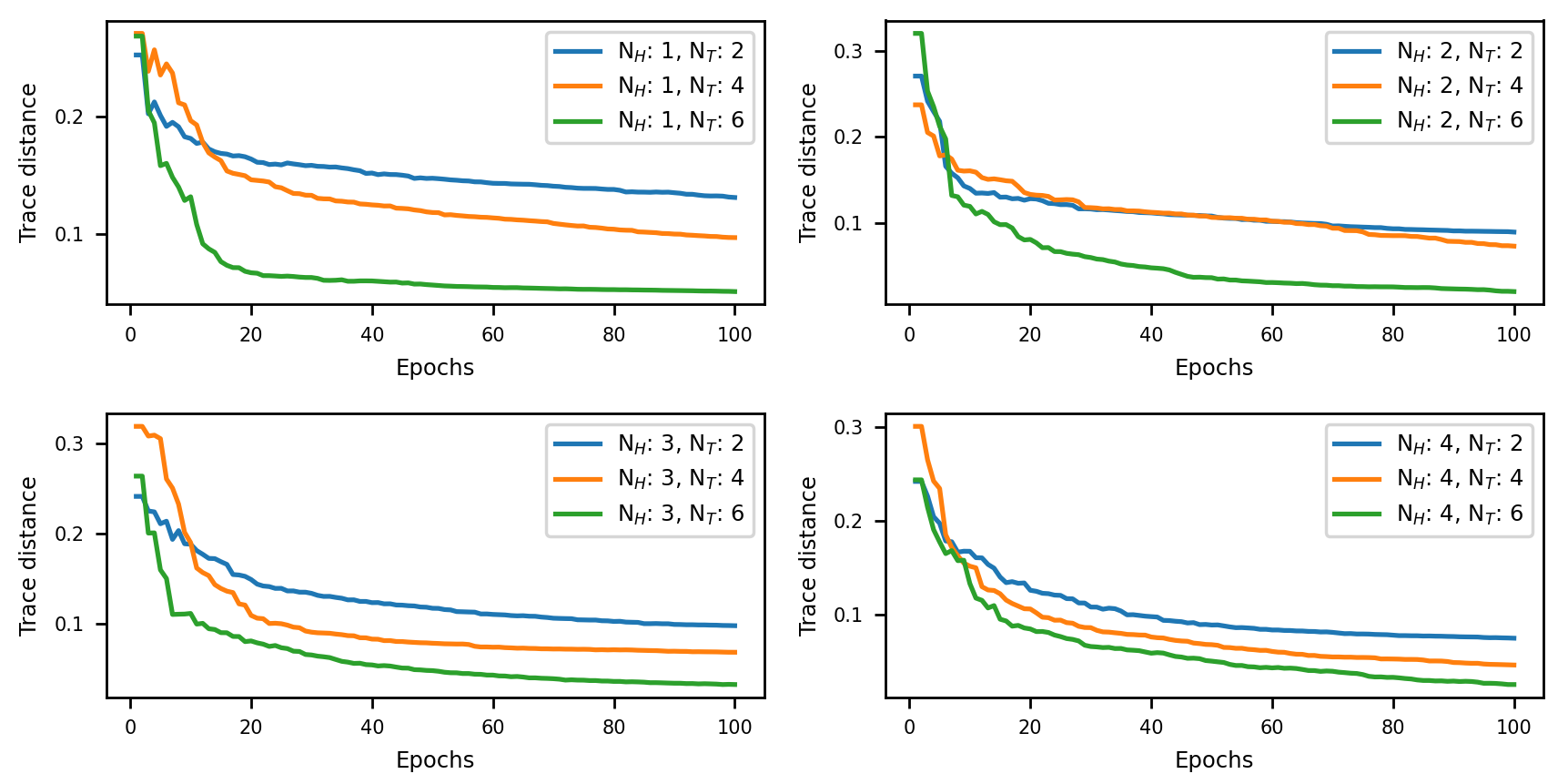} 
\caption{Trace distance between the true and the learned state is plotted as a function of the epochs during the optimization process for fixed $N_O$ = 16 and varying $N_H$, $N_T$. As is evident, using time dynamics of observables in the optimization process improves the learning of the quantum state. }
\label{fig_fstrace_same_nh}
\end{figure} 
\begin{figure}[ht!]
  \centering 
\includegraphics[width=6.2in]{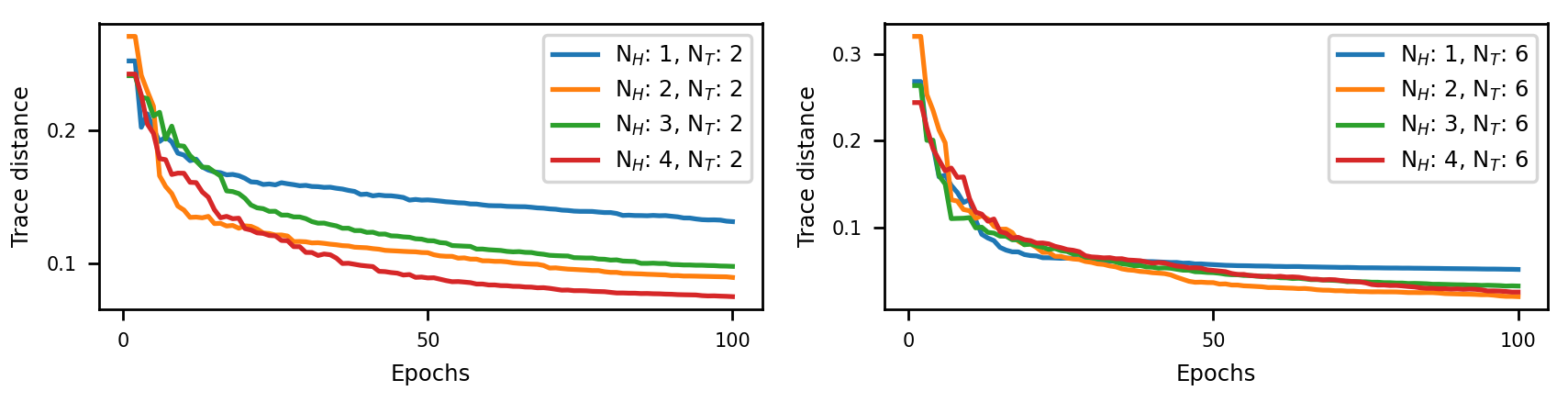} 
\caption{The effect of varying $N_H$ for fixed $N_T$ on state learning.}
\label{fig_fstrace_same_tsteps}
\end{figure} 
The operators in Eq. (\ref{basis1}) form the informationally complete (IC) Hermitian operators whose expectation values are the probabilities and coherences of a 2-qubit quantum system. We consider the time dynamics of a few of these observables in our optimization protocol as discussed in Section \ref{method}. Here, we present the results corresponding to the learning of a random 5-qubit quantum state. As was evident in the Hamiltonian learning protocol, for state learning as well, including the expectation values of observables at higher time steps in the optimization process improves the convergence of the learned state to the true state as shown in Figure \ref{fig_fstrace_same_nh}. Also, to analyze the effect of varying the number of random Hamiltonians on state learning we see in Figure \ref{fig_fstrace_same_tsteps} that as $N_H$ increases the learning improves significantly when $N_T$ is small but for higher $N_T$ the state can be learned using fewer Hamiltonians. Lastly, in Figure \ref{fig_fstrace_ops_vary} we vary the number of observables for various $N_T$ and as expected, we can learn the quantum state using fewer observables than the IC set if we include their time dynamics in the optimization protocol.

\subsection{Quantum Hamiltonian learning of the dynamics generated by the SU(3) group using analytical gradients} \label{gellmann}
The Hamiltonian of such a system can be expressed in terms of Gell-Mann matrices: $H = \sum_{i=1}^8 c_i\lambda_i$, where $\lambda_i$ are the generators of the SU(3) group:

\[
\lambda_1 = 
\begin{bmatrix}
 0 & 1 & 0  \\
  1 & 0 & 0  \\
  0 & 0 & 0
 \end{bmatrix},
 \lambda_2 = 
\begin{bmatrix}
 0 & -\textit{i} & 0  \\
  \textit{i} & 0 & 0  \\
  0 & 0 & 0
 \end{bmatrix},
 \lambda_3 = 
\begin{bmatrix}
 1 & 0 & 0  \\
  0 & -1 & 0  \\
  0 & 0 & 0
 \end{bmatrix},
 \lambda_4 = 
\begin{bmatrix}
 0 & 0 & 1  \\
  0 & 0 & 0  \\
  1 & 0 & 0
 \end{bmatrix},
\]
\[
\lambda_5 = 
\begin{bmatrix}
 0 & 0 & -\textit{i}  \\
  0 & 0 & 0  \\
  \textit{i} & 0 & 0
 \end{bmatrix},
 \lambda_6 = 
\begin{bmatrix}
 0 & 0 & 0  \\
  0 & 0 & 1  \\
  0 & 1 & 0
 \end{bmatrix},
 \lambda_7 = 
\begin{bmatrix}
 0 & 0 & 0  \\
  0 & 0 & -\textit{i}  \\
  0 & \textit{i} & 0
 \end{bmatrix},
 \lambda_8 = \frac{1}{\sqrt{3}}
\begin{bmatrix}
 1 & 0 & 0  \\
  0 & 1 & 0  \\
  0 & 0 & -2
 \end{bmatrix}.
\]

Now, in order to learn the Hamiltonian of such a quantum system we can perform optimization to learn the coefficients $\{c_i\}$. One of the ways to do so is to use our proposed formalism of using time dynamics of various observables sampled using different quantum states and optimizing the coefficients using the numerical approach of gradient calculation as discussed in Section \ref{gradient} based on the cost function defined in Eq. (\ref{cost_func}). However, as a generalized approach, we can also calculate the gradients analytically. There can be different approaches to doing so. We, hereby, propose an approach based on a version of the Baker–Campbell–Hausdorff (BCH) formula:
\begin{equation}
    e^ABe^{-A} = B + [A,B] + \frac{1}{2!}[A,[A,B]] + \frac{1}{3!}[A,[A,[A,B]] + \ldots + \frac{1}{n!}[A,[A,\ldots,[A,B]],\ldots] 
\end{equation}
The unitary operator for the state evolution under the Hamiltonian H is $U$ = $\exp{-i\sum_{i=1}^8 c_i\lambda_it}$. To compute the gradients of the cost function in Eq. (\ref{cost_func}) analytically, we use the BCH expansion of $U^\dagger O_\alpha U$ as:
\begin{eqnarray}
U^\dagger O_\alpha U = O_\alpha + \textit{i}t\left[\sum_kc_k\lambda_k,O_\alpha\right] + \frac{(\textit{i}t)^2}{2!}\left[\sum_kc_k\lambda_k,\left[\sum_{k^\prime} c_{k^\prime}\lambda_{k^\prime},O_\alpha\right]\right] + \ldots \nonumber \\
+ \hspace{0.2cm} \frac{(\textit{i}t)^n}{n!}\left[\sum_kc_k\lambda_k,\left[\sum_{k^\prime}c_{k^\prime}\lambda_{k^\prime},\ldots,\left[\sum_{k^{n^\prime}}c_{k^{n^\prime}}\lambda_{k^{n^\prime}},O_\alpha\right]\right],\ldots,\right]
\end{eqnarray}
Now, we take the derivative of $U^\dagger O_\alpha U$ with respect to the coefficients $\{c_p\}$:
\begin{eqnarray}
    \frac{\partial U^\dagger O_\alpha U}{\partial c_p} = \textit{i}t\left[\lambda_p,O_\alpha\right] + \frac{(\textit{i}t)^2}{2!}\left(\left[\sum_{k^\prime \neq p}c_{k^\prime}\lambda_p,[\lambda_{k^\prime},O_\alpha]\right] + 2c_p[\lambda_p,[\lambda_p,O_\alpha]]\right) + \ldots \nonumber \\
    + \hspace{0.1cm} \frac{(\textit{i}t)^n}{n!}(\left[\sum_{k^\prime \neq p}\sum_{k^{\prime\prime} \neq p}\ldots\sum_{k^{n\prime} \neq p} c_{k^\prime}c_{k^{\prime\prime}}\ldots c_{k^{n\prime}} \lambda_p[\lambda_{k^\prime},[\lambda_{k^{\prime\prime}},\ldots,[\lambda_{k^{n^\prime}},O_\alpha]]\ldots\right] \nonumber \\ + \left[\sum_{k^{\prime\prime} \neq p}\sum_{k^{\prime\prime\prime} \neq p}\ldots\sum_{k^{n\prime} \neq p} 2c_p c_{k^{\prime\prime}}c_{k^{\prime\prime\prime}}\ldots c_{k^{n\prime}} \lambda_p,[\lambda_p,[\lambda_k^{\prime\prime},\ldots,[\lambda_{k^{n\prime}},O_\alpha]],\ldots\right] \nonumber \\
    + \hspace{0.2cm} [nc_p^{n-1}\lambda_p,[\lambda_p,\ldots[\lambda_p,O_\alpha]]\ldots]) \label{grad}
\end{eqnarray}
The Gell-Mann matrices satisfy the commutation relationship:
\begin{equation}
    [\lambda_a,\lambda_b] = 2\textit{i}\sum_c f^{abc}\lambda_c
\end{equation}
where $f^{abc}$ are the structure constants given by: $f^{abc} = -\frac{1}{4}i( tr(\lambda_a[\lambda_b,\lambda_c]))$. Thus, depending on the required accuracy level, one can fix the value of \textit{n} and calculate $\frac{\partial U^\dagger O_\alpha U}{\partial c_p}$. Thereby, we can calculate the gradients as:
\begin{equation}
    \frac{\partial Cost}{\partial c_p} = 2 \sum_{\alpha ,i,k}^{N_O, N_S, N_T}\bra{\psi_i}\frac{\partial U^\dagger(k\Delta t) O_\alpha U(k\Delta t)}{\partial c_p}\ket{\psi_i} \label{Full_grad_SU3}
\end{equation}
Thus, substituting Eq. (\ref{grad}) in Eq. (\ref{Full_grad_SU3}) and taking advantage of the structural relationship of SU(N) systems, the gradients can be computed with measurements that are independent of the time steps of the evolution. This protocol can be extended to SU(N) systems as well and we intend to investigate it further in future works.

% \section*{Data Availability}
% The data generated during the current study is available from the corresponding author upon reasonable request.

% %%%%%%%%%%%%%%%%%%%%%%%%%%%%%%%%%%%%%%%%%%%%%%%%%%%%%%%%%%%%%%%%%%%%%
% %% The "Acknowledgement" section can be given in all manuscript
% %% classes.  This should be given within the "acknowledgment"
% %% environment, which will make the correct section or running title.
% %%%%%%%%%%%%%%%%%%%%%%%%%%%%%%%%%%%%%%%%%%%%%%%%%%%%%%%%%%%%%%%%%%%%%
\begin{acknowledgement}
We acknowledge the financial support from the National Science Foundation under Award No. 1955907. We also like to acknowledge the partial financial support by the U.S. Department of Energy (Office of Basic Energy Sciences) under Award No. DE-SC0019215. We also acknowledge the use of the IBM Q for this work. The views expressed here are those of the authors and do not reflect the official policy or position of IBM or the IBM Q team.

\end{acknowledgement}

% \section*{Competing Interests}
% The Authors declare no Competing Financial or Non-Financial Interests.

% \begin{suppinfo}

% This will usually read something like: ``Experimental procedures and
% characterization data for all new compounds. The class will
% automatically add a sentence pointing to the information on-line:

% \end{suppinfo}

% \bibliographystyle{unsrt}
\bibliography{achemso-demo}

\providecommand{\latin}[1]{#1}
\makeatletter
\providecommand{\doi}
  {\begingroup\let\do\@makeother\dospecials
  \catcode`\{=1 \catcode`\}=2 \doi@aux}
\providecommand{\doi@aux}[1]{\endgroup\texttt{#1}}
\makeatother
\providecommand*\mcitethebibliography{\thebibliography}
\csname @ifundefined\endcsname{endmcitethebibliography}
  {\let\endmcitethebibliography\endthebibliography}{}
\begin{mcitethebibliography}{52}
\providecommand*\natexlab[1]{#1}
\providecommand*\mciteSetBstSublistMode[1]{}
\providecommand*\mciteSetBstMaxWidthForm[2]{}
\providecommand*\mciteBstWouldAddEndPuncttrue
  {\def\EndOfBibitem{\unskip.}}
\providecommand*\mciteBstWouldAddEndPunctfalse
  {\let\EndOfBibitem\relax}
\providecommand*\mciteSetBstMidEndSepPunct[3]{}
\providecommand*\mciteSetBstSublistLabelBeginEnd[3]{}
\providecommand*\EndOfBibitem{}
\mciteSetBstSublistMode{f}
\mciteSetBstMaxWidthForm{subitem}{(\alph{mcitesubitemcount})}
\mciteSetBstSublistLabelBeginEnd
  {\mcitemaxwidthsubitemform\space}
  {\relax}
  {\relax}

\bibitem[Nielsen and Chuang(2011)Nielsen, and Chuang]{nielson}
Nielsen,~M.~A.; Chuang,~I.~L. \emph{Quantum Computation and Quantum
  Information: 10th Anniversary Edition}, 10th ed.; Cambridge University Press:
  USA, 2011\relax
\mciteBstWouldAddEndPuncttrue
\mciteSetBstMidEndSepPunct{\mcitedefaultmidpunct}
{\mcitedefaultendpunct}{\mcitedefaultseppunct}\relax
\EndOfBibitem
\bibitem[Kais(2014)]{kais}
Kais,~S. \emph{Quantum Information and Computation for Chemistry}; Wiley and
  Sons: Hoboken: NJ, 2014; Vol. 154\relax
\mciteBstWouldAddEndPuncttrue
\mciteSetBstMidEndSepPunct{\mcitedefaultmidpunct}
{\mcitedefaultendpunct}{\mcitedefaultseppunct}\relax
\EndOfBibitem
\bibitem[James \latin{et~al.}(2001)James, Kwiat, Munro, and White]{qubits}
James,~D. F.~V.; Kwiat,~P.~G.; Munro,~W.~J.; White,~A.~G. Measurement of
  qubits. \emph{Phys. Rev. A} \textbf{2001}, \emph{64}, 052312\relax
\mciteBstWouldAddEndPuncttrue
\mciteSetBstMidEndSepPunct{\mcitedefaultmidpunct}
{\mcitedefaultendpunct}{\mcitedefaultseppunct}\relax
\EndOfBibitem
\bibitem[Banaszek \latin{et~al.}(2013)Banaszek, Cramer, and Gross]{cramer}
Banaszek,~K.; Cramer,~M.; Gross,~D. Focus on quantum tomography. \emph{New J.
  Phys.} \textbf{2013}, \emph{15}, 125020\relax
\mciteBstWouldAddEndPuncttrue
\mciteSetBstMidEndSepPunct{\mcitedefaultmidpunct}
{\mcitedefaultendpunct}{\mcitedefaultseppunct}\relax
\EndOfBibitem
\bibitem[Paris and Rehacek(2004)Paris, and Rehacek]{paris2004quantum}
Paris,~M.; Rehacek,~J. \emph{Quantum state estimation}; Springer Science \&
  Business Media, 2004; Vol. 649\relax
\mciteBstWouldAddEndPuncttrue
\mciteSetBstMidEndSepPunct{\mcitedefaultmidpunct}
{\mcitedefaultendpunct}{\mcitedefaultseppunct}\relax
\EndOfBibitem
\bibitem[Artiles \latin{et~al.}(2005)Artiles, Gill, and
  Gut{\c{}}~{\u{a}}]{artiles2005invitation}
Artiles,~L.~M.; Gill,~R.~D.; Gut{\c{}}~{\u{a}},~M. An invitation to quantum
  tomography. \emph{Journal of the Royal Statistical Society: Series B
  (Statistical Methodology)} \textbf{2005}, \emph{67}, 109--134\relax
\mciteBstWouldAddEndPuncttrue
\mciteSetBstMidEndSepPunct{\mcitedefaultmidpunct}
{\mcitedefaultendpunct}{\mcitedefaultseppunct}\relax
\EndOfBibitem
\bibitem[Yu \latin{et~al.}(2022)Yu, Sun, Han, and Yuan]{yu2022practical}
Yu,~W.; Sun,~J.; Han,~Z.; Yuan,~X. Practical and Efficient Hamiltonian
  Learning. \emph{arXiv preprint arXiv:2201.00190} \textbf{2022}, \relax
\mciteBstWouldAddEndPunctfalse
\mciteSetBstMidEndSepPunct{\mcitedefaultmidpunct}
{}{\mcitedefaultseppunct}\relax
\EndOfBibitem
\bibitem[Haah \latin{et~al.}(2021)Haah, Kothari, and Tang]{haah2021optimal}
Haah,~J.; Kothari,~R.; Tang,~E. Optimal learning of quantum Hamiltonians from
  high-temperature Gibbs states. \emph{arXiv preprint arXiv:2108.04842}
  \textbf{2021}, \relax
\mciteBstWouldAddEndPunctfalse
\mciteSetBstMidEndSepPunct{\mcitedefaultmidpunct}
{}{\mcitedefaultseppunct}\relax
\EndOfBibitem
\bibitem[Krastanov \latin{et~al.}(2019)Krastanov, Zhou, Flammia, and
  Jiang]{krastanov2019stochastic}
Krastanov,~S.; Zhou,~S.; Flammia,~S.~T.; Jiang,~L. Stochastic estimation of
  dynamical variables. \emph{Quantum Science and Technology} \textbf{2019},
  \emph{4}, 035003\relax
\mciteBstWouldAddEndPuncttrue
\mciteSetBstMidEndSepPunct{\mcitedefaultmidpunct}
{\mcitedefaultendpunct}{\mcitedefaultseppunct}\relax
\EndOfBibitem
\bibitem[Evans \latin{et~al.}(2019)Evans, Harper, and
  Flammia]{evans2019scalable}
Evans,~T.~J.; Harper,~R.; Flammia,~S.~T. Scalable bayesian hamiltonian
  learning. \emph{arXiv preprint arXiv:1912.07636} \textbf{2019}, \relax
\mciteBstWouldAddEndPunctfalse
\mciteSetBstMidEndSepPunct{\mcitedefaultmidpunct}
{}{\mcitedefaultseppunct}\relax
\EndOfBibitem
\bibitem[Bairey \latin{et~al.}(2019)Bairey, Arad, and
  Lindner]{bairey2019learning}
Bairey,~E.; Arad,~I.; Lindner,~N.~H. Learning a local Hamiltonian from local
  measurements. \emph{Physical review letters} \textbf{2019}, \emph{122},
  020504\relax
\mciteBstWouldAddEndPuncttrue
\mciteSetBstMidEndSepPunct{\mcitedefaultmidpunct}
{\mcitedefaultendpunct}{\mcitedefaultseppunct}\relax
\EndOfBibitem
\bibitem[Qi and Ranard(2019)Qi, and Ranard]{qi2019determining}
Qi,~X.-L.; Ranard,~D. Determining a local Hamiltonian from a single eigenstate.
  \emph{Quantum} \textbf{2019}, \emph{3}, 159\relax
\mciteBstWouldAddEndPuncttrue
\mciteSetBstMidEndSepPunct{\mcitedefaultmidpunct}
{\mcitedefaultendpunct}{\mcitedefaultseppunct}\relax
\EndOfBibitem
\bibitem[Poyatos \latin{et~al.}(1997)Poyatos, Cirac, and
  Zoller]{PhysRevLett.78.390}
Poyatos,~J.~F.; Cirac,~J.~I.; Zoller,~P. Complete Characterization of a Quantum
  Process: The Two-Bit Quantum Gate. \emph{Phys. Rev. Lett.} \textbf{1997},
  \emph{78}, 390--393\relax
\mciteBstWouldAddEndPuncttrue
\mciteSetBstMidEndSepPunct{\mcitedefaultmidpunct}
{\mcitedefaultendpunct}{\mcitedefaultseppunct}\relax
\EndOfBibitem
\bibitem[D'Ariano and Lo~Presti(2001)D'Ariano, and
  Lo~Presti]{PhysRevLett.86.4195}
D'Ariano,~G.~M.; Lo~Presti,~P. Quantum Tomography for Measuring Experimentally
  the Matrix Elements of an Arbitrary Quantum Operation. \emph{Phys. Rev.
  Lett.} \textbf{2001}, \emph{86}, 4195--4198\relax
\mciteBstWouldAddEndPuncttrue
\mciteSetBstMidEndSepPunct{\mcitedefaultmidpunct}
{\mcitedefaultendpunct}{\mcitedefaultseppunct}\relax
\EndOfBibitem
\bibitem[Altepeter \latin{et~al.}(2003)Altepeter, Branning, Jeffrey, Wei,
  Kwiat, Thew, O'Brien, Nielsen, and White]{PhysRevLett.90.193601}
Altepeter,~J.~B.; Branning,~D.; Jeffrey,~E.; Wei,~T.~C.; Kwiat,~P.~G.;
  Thew,~R.~T.; O'Brien,~J.~L.; Nielsen,~M.~A.; White,~A.~G. Ancilla-Assisted
  Quantum Process Tomography. \emph{Phys. Rev. Lett.} \textbf{2003}, \emph{90},
  193601\relax
\mciteBstWouldAddEndPuncttrue
\mciteSetBstMidEndSepPunct{\mcitedefaultmidpunct}
{\mcitedefaultendpunct}{\mcitedefaultseppunct}\relax
\EndOfBibitem
\bibitem[D'Ariano and Lo~Presti(2003)D'Ariano, and
  Lo~Presti]{PhysRevLett.91.047902}
D'Ariano,~G.~M.; Lo~Presti,~P. Imprinting Complete Information about a Quantum
  Channel on its Output State. \emph{Phys. Rev. Lett.} \textbf{2003},
  \emph{91}, 047902\relax
\mciteBstWouldAddEndPuncttrue
\mciteSetBstMidEndSepPunct{\mcitedefaultmidpunct}
{\mcitedefaultendpunct}{\mcitedefaultseppunct}\relax
\EndOfBibitem
\bibitem[Ekert \latin{et~al.}(2002)Ekert, Alves, Oi, Horodecki, Horodecki, and
  Kwek]{PhysRevLett.88.217901}
Ekert,~A.~K.; Alves,~C.~M.; Oi,~D. K.~L.; Horodecki,~M.; Horodecki,~P.;
  Kwek,~L.~C. Direct Estimations of Linear and Nonlinear Functionals of a
  Quantum State. \emph{Phys. Rev. Lett.} \textbf{2002}, \emph{88}, 217901\relax
\mciteBstWouldAddEndPuncttrue
\mciteSetBstMidEndSepPunct{\mcitedefaultmidpunct}
{\mcitedefaultendpunct}{\mcitedefaultseppunct}\relax
\EndOfBibitem
\bibitem[Horodecki and Ekert(2002)Horodecki, and Ekert]{PhysRevLett.89.127902}
Horodecki,~P.; Ekert,~A. Method for Direct Detection of Quantum Entanglement.
  \emph{Phys. Rev. Lett.} \textbf{2002}, \emph{89}, 127902\relax
\mciteBstWouldAddEndPuncttrue
\mciteSetBstMidEndSepPunct{\mcitedefaultmidpunct}
{\mcitedefaultendpunct}{\mcitedefaultseppunct}\relax
\EndOfBibitem
\bibitem[Bovino \latin{et~al.}(2005)Bovino, Castagnoli, Ekert, Horodecki,
  Alves, and Sergienko]{PhysRevLett.95.240407}
Bovino,~F.~A.; Castagnoli,~G.; Ekert,~A.; Horodecki,~P.; Alves,~C.~M.;
  Sergienko,~A.~V. Direct Measurement of Nonlinear Properties of Bipartite
  Quantum States. \emph{Phys. Rev. Lett.} \textbf{2005}, \emph{95},
  240407\relax
\mciteBstWouldAddEndPuncttrue
\mciteSetBstMidEndSepPunct{\mcitedefaultmidpunct}
{\mcitedefaultendpunct}{\mcitedefaultseppunct}\relax
\EndOfBibitem
\bibitem[T\'oth \latin{et~al.}(2010)T\'oth, Wieczorek, Gross, Krischek,
  Schwemmer, and Weinfurter]{PhysRevLett.105.250403}
T\'oth,~G.; Wieczorek,~W.; Gross,~D.; Krischek,~R.; Schwemmer,~C.;
  Weinfurter,~H. Permutationally Invariant Quantum Tomography. \emph{Phys. Rev.
  Lett.} \textbf{2010}, \emph{105}, 250403\relax
\mciteBstWouldAddEndPuncttrue
\mciteSetBstMidEndSepPunct{\mcitedefaultmidpunct}
{\mcitedefaultendpunct}{\mcitedefaultseppunct}\relax
\EndOfBibitem
\bibitem[Gupta \latin{et~al.}(2021)Gupta, Levine, and
  Kais]{gupta2021convergence}
Gupta,~R.; Levine,~R.~D.; Kais,~S. Convergence of a Reconstructed Density
  Matrix to a Pure State Using the Maximal Entropy Approach. \emph{The Journal
  of Physical Chemistry A} \textbf{2021}, \emph{125}, 7588--7595\relax
\mciteBstWouldAddEndPuncttrue
\mciteSetBstMidEndSepPunct{\mcitedefaultmidpunct}
{\mcitedefaultendpunct}{\mcitedefaultseppunct}\relax
\EndOfBibitem
\bibitem[Gupta \latin{et~al.}(2021)Gupta, Xia, Levine, and
  Kais]{gupta2021maximal}
Gupta,~R.; Xia,~R.; Levine,~R.~D.; Kais,~S. Maximal entropy approach for
  quantum state tomography. \emph{PRX Quantum} \textbf{2021}, \emph{2},
  010318\relax
\mciteBstWouldAddEndPuncttrue
\mciteSetBstMidEndSepPunct{\mcitedefaultmidpunct}
{\mcitedefaultendpunct}{\mcitedefaultseppunct}\relax
\EndOfBibitem
\bibitem[Gupta \latin{et~al.}(2022)Gupta, Sajjan, Levine, and
  Kais]{gupta2022variation}
Gupta,~R.; Sajjan,~M.; Levine,~R.~D.; Kais,~S. Variational approach to quantum
  state tomography based on maximal entropy formalism. \emph{Phys. Chem. Chem.
  Phys.} \textbf{2022}, \emph{24}, 28870--28877\relax
\mciteBstWouldAddEndPuncttrue
\mciteSetBstMidEndSepPunct{\mcitedefaultmidpunct}
{\mcitedefaultendpunct}{\mcitedefaultseppunct}\relax
\EndOfBibitem
\bibitem[Aaronson(2017)]{aransonst}
Aaronson,~S. Shadow Tomography of Quantum States. 2017;
  \url{https://arxiv.org/abs/1711.01053}\relax
\mciteBstWouldAddEndPuncttrue
\mciteSetBstMidEndSepPunct{\mcitedefaultmidpunct}
{\mcitedefaultendpunct}{\mcitedefaultseppunct}\relax
\EndOfBibitem
\bibitem[Chen \latin{et~al.}(2022)Chen, Huang, Li, Liu, and
  Sellke]{chen2022tight}
Chen,~S.; Huang,~B.; Li,~J.; Liu,~A.; Sellke,~M. Tight bounds for state
  tomography with incoherent measurements. \emph{arXiv preprint
  arXiv:2206.05265} \textbf{2022}, \relax
\mciteBstWouldAddEndPunctfalse
\mciteSetBstMidEndSepPunct{\mcitedefaultmidpunct}
{}{\mcitedefaultseppunct}\relax
\EndOfBibitem
\bibitem[Huang \latin{et~al.}(2020)Huang, Kueng, and
  Preskill]{huang2020predicting}
Huang,~H.-Y.; Kueng,~R.; Preskill,~J. Predicting many properties of a quantum
  system from very few measurements. \emph{Nature Physics} \textbf{2020},
  \emph{16}, 1050--1057\relax
\mciteBstWouldAddEndPuncttrue
\mciteSetBstMidEndSepPunct{\mcitedefaultmidpunct}
{\mcitedefaultendpunct}{\mcitedefaultseppunct}\relax
\EndOfBibitem
\bibitem[Schuch \latin{et~al.}(2008)Schuch, Cirac, and
  Verstraete]{schuch2008computational}
Schuch,~N.; Cirac,~I.; Verstraete,~F. Computational difficulty of finding
  matrix product ground states. \emph{Physical review letters} \textbf{2008},
  \emph{100}, 250501\relax
\mciteBstWouldAddEndPuncttrue
\mciteSetBstMidEndSepPunct{\mcitedefaultmidpunct}
{\mcitedefaultendpunct}{\mcitedefaultseppunct}\relax
\EndOfBibitem
\bibitem[Bilgin and Boixo(2010)Bilgin, and Boixo]{bilgin2010preparing}
Bilgin,~E.; Boixo,~S. Preparing thermal states of quantum systems by dimension
  reduction. \emph{Physical review letters} \textbf{2010}, \emph{105},
  170405\relax
\mciteBstWouldAddEndPuncttrue
\mciteSetBstMidEndSepPunct{\mcitedefaultmidpunct}
{\mcitedefaultendpunct}{\mcitedefaultseppunct}\relax
\EndOfBibitem
\bibitem[McClean \latin{et~al.}(2016)McClean, Romero, Babbush, and
  Aspuru-Guzik]{mcclean2016theory}
McClean,~J.~R.; Romero,~J.; Babbush,~R.; Aspuru-Guzik,~A. The theory of
  variational hybrid quantum-classical algorithms. \emph{New Journal of
  Physics} \textbf{2016}, \emph{18}, 023023\relax
\mciteBstWouldAddEndPuncttrue
\mciteSetBstMidEndSepPunct{\mcitedefaultmidpunct}
{\mcitedefaultendpunct}{\mcitedefaultseppunct}\relax
\EndOfBibitem
\bibitem[Bravo-Prieto \latin{et~al.}(2019)Bravo-Prieto, LaRose, Cerezo, Subasi,
  Cincio, and Coles]{bravo2019variational}
Bravo-Prieto,~C.; LaRose,~R.; Cerezo,~M.; Subasi,~Y.; Cincio,~L.; Coles,~P.~J.
  Variational quantum linear solver. \emph{arXiv preprint arXiv:1909.05820}
  \textbf{2019}, \relax
\mciteBstWouldAddEndPunctfalse
\mciteSetBstMidEndSepPunct{\mcitedefaultmidpunct}
{}{\mcitedefaultseppunct}\relax
\EndOfBibitem
\bibitem[Huang \latin{et~al.}(2019)Huang, Bharti, and
  Rebentrost]{huang2019near}
Huang,~H.-Y.; Bharti,~K.; Rebentrost,~P. Near-term quantum algorithms for
  linear systems of equations. \emph{arXiv preprint arXiv:1909.07344}
  \textbf{2019}, \relax
\mciteBstWouldAddEndPunctfalse
\mciteSetBstMidEndSepPunct{\mcitedefaultmidpunct}
{}{\mcitedefaultseppunct}\relax
\EndOfBibitem
\bibitem[LaRose \latin{et~al.}(2019)LaRose, Tikku, O’Neel-Judy, Cincio, and
  Coles]{larose2019variational}
LaRose,~R.; Tikku,~A.; O’Neel-Judy,~{\'E}.; Cincio,~L.; Coles,~P.~J.
  Variational quantum state diagonalization. \emph{npj Quantum Information}
  \textbf{2019}, \emph{5}, 1--10\relax
\mciteBstWouldAddEndPuncttrue
\mciteSetBstMidEndSepPunct{\mcitedefaultmidpunct}
{\mcitedefaultendpunct}{\mcitedefaultseppunct}\relax
\EndOfBibitem
\bibitem[Cerezo \latin{et~al.}(2020)Cerezo, Sharma, Arrasmith, and
  Coles]{cerezo2020variational}
Cerezo,~M.; Sharma,~K.; Arrasmith,~A.; Coles,~P.~J. Variational quantum state
  eigensolver. \emph{arXiv preprint arXiv:2004.01372} \textbf{2020}, \relax
\mciteBstWouldAddEndPunctfalse
\mciteSetBstMidEndSepPunct{\mcitedefaultmidpunct}
{}{\mcitedefaultseppunct}\relax
\EndOfBibitem
\bibitem[Peruzzo \latin{et~al.}(2014)Peruzzo, McClean, Shadbolt, Yung, Zhou,
  Love, Aspuru-Guzik, and O’brien]{peruzzo2014variational}
Peruzzo,~A.; McClean,~J.; Shadbolt,~P.; Yung,~M.-H.; Zhou,~X.-Q.; Love,~P.~J.;
  Aspuru-Guzik,~A.; O’brien,~J.~L. A variational eigenvalue solver on a
  photonic quantum processor. \emph{Nature communications} \textbf{2014},
  \emph{5}, 1--7\relax
\mciteBstWouldAddEndPuncttrue
\mciteSetBstMidEndSepPunct{\mcitedefaultmidpunct}
{\mcitedefaultendpunct}{\mcitedefaultseppunct}\relax
\EndOfBibitem
\bibitem[Chen \latin{et~al.}(2021)Chen, Song, Zhao, and
  Wang]{chen2021variational}
Chen,~R.; Song,~Z.; Zhao,~X.; Wang,~X. Variational quantum algorithms for trace
  distance and fidelity estimation. \emph{Quantum Science and Technology}
  \textbf{2021}, \emph{7}, 015019\relax
\mciteBstWouldAddEndPuncttrue
\mciteSetBstMidEndSepPunct{\mcitedefaultmidpunct}
{\mcitedefaultendpunct}{\mcitedefaultseppunct}\relax
\EndOfBibitem
\bibitem[Sureshbabu \latin{et~al.}(2021)Sureshbabu, Sajjan, Oh, and
  Kais]{sureshbabu2021implementation}
Sureshbabu,~S.~H.; Sajjan,~M.; Oh,~S.; Kais,~S. Implementation of Quantum
  Machine Learning for Electronic Structure Calculations of Periodic Systems on
  Quantum Computing Devices. \emph{Journal of Chemical Information and
  Modeling} \textbf{2021}, \relax
\mciteBstWouldAddEndPunctfalse
\mciteSetBstMidEndSepPunct{\mcitedefaultmidpunct}
{}{\mcitedefaultseppunct}\relax
\EndOfBibitem
\bibitem[Sajjan \latin{et~al.}(2021)Sajjan, Sureshbabu, and
  Kais]{sajjan2021quantum}
Sajjan,~M.; Sureshbabu,~S.~H.; Kais,~S. Quantum machine-learning for eigenstate
  filtration in two-dimensional materials. \emph{Journal of the American
  Chemical Society} \textbf{2021}, \emph{143}, 18426--18445\relax
\mciteBstWouldAddEndPuncttrue
\mciteSetBstMidEndSepPunct{\mcitedefaultmidpunct}
{\mcitedefaultendpunct}{\mcitedefaultseppunct}\relax
\EndOfBibitem
\bibitem[Selvarajan \latin{et~al.}(2022)Selvarajan, Sajjan, and
  Kais]{selvarajan2022variational}
Selvarajan,~R.; Sajjan,~M.; Kais,~S. Variational quantum circuits to prepare
  low energy symmetry states. \emph{Symmetry} \textbf{2022}, \emph{14},
  457\relax
\mciteBstWouldAddEndPuncttrue
\mciteSetBstMidEndSepPunct{\mcitedefaultmidpunct}
{\mcitedefaultendpunct}{\mcitedefaultseppunct}\relax
\EndOfBibitem
\bibitem[Dixit \latin{et~al.}(2021)Dixit, Selvarajan, Aldwairi, Koshka,
  Novotny, Humble, Alam, and Kais]{dixit2021training}
Dixit,~V.; Selvarajan,~R.; Aldwairi,~T.; Koshka,~Y.; Novotny,~M.~A.;
  Humble,~T.~S.; Alam,~M.~A.; Kais,~S. Training a quantum annealing based
  restricted boltzmann machine on cybersecurity data. \emph{IEEE Transactions
  on Emerging Topics in Computational Intelligence} \textbf{2021}, \emph{6},
  417--428\relax
\mciteBstWouldAddEndPuncttrue
\mciteSetBstMidEndSepPunct{\mcitedefaultmidpunct}
{\mcitedefaultendpunct}{\mcitedefaultseppunct}\relax
\EndOfBibitem
\bibitem[Sajjan \latin{et~al.}(2022)Sajjan, Li, Selvarajan, Sureshbabu, Kale,
  Gupta, Singh, and Kais]{D2CS00203E}
Sajjan,~M.; Li,~J.; Selvarajan,~R.; Sureshbabu,~S.~H.; Kale,~S.~S.; Gupta,~R.;
  Singh,~V.; Kais,~S. Quantum machine learning for chemistry and physics.
  \emph{Chem. Soc. Rev.} \textbf{2022}, \emph{51}, 6475--6573\relax
\mciteBstWouldAddEndPuncttrue
\mciteSetBstMidEndSepPunct{\mcitedefaultmidpunct}
{\mcitedefaultendpunct}{\mcitedefaultseppunct}\relax
\EndOfBibitem
\bibitem[Cao \latin{et~al.}(2019)Cao, Romero, Olson, Degroote, Johnson,
  Kieferov{\'a}, Kivlichan, Menke, Peropadre, Sawaya, \latin{et~al.}
  others]{cao2019quantum}
Cao,~Y.; Romero,~J.; Olson,~J.~P.; Degroote,~M.; Johnson,~P.~D.;
  Kieferov{\'a},~M.; Kivlichan,~I.~D.; Menke,~T.; Peropadre,~B.; Sawaya,~N.~P.,
  \latin{et~al.}  Quantum chemistry in the age of quantum computing.
  \emph{Chemical reviews} \textbf{2019}, \emph{119}, 10856--10915\relax
\mciteBstWouldAddEndPuncttrue
\mciteSetBstMidEndSepPunct{\mcitedefaultmidpunct}
{\mcitedefaultendpunct}{\mcitedefaultseppunct}\relax
\EndOfBibitem
\bibitem[Crooks(2019)]{crooks2019gradients}
Crooks,~G.~E. Gradients of parameterized quantum gates using the
  parameter-shift rule and gate decomposition. \emph{arXiv preprint
  arXiv:1905.13311} \textbf{2019}, \relax
\mciteBstWouldAddEndPunctfalse
\mciteSetBstMidEndSepPunct{\mcitedefaultmidpunct}
{}{\mcitedefaultseppunct}\relax
\EndOfBibitem
\bibitem[Selvarajan \latin{et~al.}(2021)Selvarajan, Dixit, Cui, Humble, and
  Kais]{selvarajan2021prime}
Selvarajan,~R.; Dixit,~V.; Cui,~X.; Humble,~T.~S.; Kais,~S. Prime factorization
  using quantum variational imaginary time evolution. \emph{Scientific reports}
  \textbf{2021}, \emph{11}, 1--8\relax
\mciteBstWouldAddEndPuncttrue
\mciteSetBstMidEndSepPunct{\mcitedefaultmidpunct}
{\mcitedefaultendpunct}{\mcitedefaultseppunct}\relax
\EndOfBibitem
\bibitem[Abraham \latin{et~al.}(2019)Abraham, AduOffei, Agarwal, Akhalwaya,
  Aleksandrowicz, Alexander, Arbel, Asfaw, Azaustre, AzizNgoueya, Bansal,
  Barkoutsos, Barron, Bello, Ben-Haim, Bevenius, Bishop, Bolos, Bosch, Bravyi,
  Bucher, Burov, Cabrera, Calpin, Capelluto, Carballo, Carrascal, Chen, Chen,
  Chen, Chen, Chen, Chow, Churchill, Claus, Clauss, Cocking, Cross, Cross,
  Cross, Cruz-Benito, Culver, C{\'o}rcoles-Gonzales, Dague, Dandachi, Daniels,
  Dartiailh, DavideFrr, Davila, Dekusar, Ding, Doi, Drechsler, Drew,
  Dumitrescu, Dumon, Duran, EL-Safty, Eastman, Eendebak, Egger, Everitt,
  Fern{\'a}ndez, Ferrera, Fouilland, FranckChevallier, Frisch, Fuhrer, GEORGE,
  Gacon, Gago, Gambella, Gambetta, Gammanpila, Garcia, Garion, Gilliam,
  Giridharan, Gomez-Mosquera, de~la Puente~Gonz{\'a}lez, Gorzinski, Gould,
  Greenberg, Grinko, Guan, Gunnels, Haglund, Haide, Hamamura, Hamido, Havlicek,
  Hellmers, Herok, Hillmich, Horii, Howington, Hu, Hu, Huisman, Imai, Imamichi,
  Ishizaki, Iten, Itoko, JamesSeaward, Javadi, Javadi-Abhari, Jessica,
  Jivrajani, Johns, Jonathan-Shoemaker, Kachmann, Kanazawa, Kang-Bae, Karazeev,
  Kassebaum, King, Knabberjoe, Kobayashi, Kovyrshin, Krishnakumar, Krishnan,
  Krsulich, Kus, LaRose, Lacal, Lambert, Lapeyre, Latone, Lawrence, Lee, Li,
  Liu, Liu, Maeng, Malyshev, Manela, Marecek, Marques, Maslov, Mathews, Matsuo,
  McClure, McGarry, McKay, McPherson, Meesala, Metcalfe, Mevissen, Mezzacapo,
  Midha, Minev, Mitchell, Moll, Mooring, Morales, Moran, MrF, Murali,
  M{\"u}ggenburg, Nadlinger, Nakanishi, Nannicini, Nation, Navarro, Naveh,
  Neagle, Neuweiler, Niroula, Norlen, O'Riordan, Ogunbayo, Ollitrault, Oud,
  Padilha, Paik, Pang, Perriello, Phan, Piro, Pistoia, Piveteau,
  Pozas-iKerstjens, Prutyanov, Puzzuoli, P{\'e}rez, Quintiii, Rahman, Raja,
  Ramagiri, Rao, Raymond, Redondo, Reuter, Rice, Rodr{\'\i}guez, RohithKarur,
  Rossmannek, Ryu, SAPV, SamFerracin, Sandberg, Sapra, Sargsyan, Sarkar,
  Sathaye, Schmitt, Schnabel, Schoenfeld, Scholten, Schoute, Schwarm, Sertage,
  Setia, Shammah, Shi, Silva, Simonetto, Singstock, Siraichi, Sitdikov,
  Sivarajah, Sletfjerding, Smolin, Soeken, Sokolov, SooluThomas, Starfish,
  Steenken, Stypulkoski, Sun, Sung, Takahashi, Tavernelli, Taylor, Taylour,
  Thomas, Tillet, Tod, Tomasik, de~la Torre, Trabing, Treinish, TrishaPe,
  Turner, Vaknin, Valcarce, Varchon, Vazquez, Villar, Vogt-Lee, Vuillot,
  Weaver, Wieczorek, Wildstrom, Winston, Woehr, Woerner, Woo, Wood, Wood, Wood,
  Wood, Wootton, Yeralin, Yonge-Mallo, Young, Yu, Zachow, Zdanski, Zhang,
  Zoufal, Zoufalc, a~kapila, a~matsuo, bcamorrison, brandhsn, chlorophyll zz,
  dekel.meirom, dekool, dime10, drholmie, dtrenev, ehchen, elfrocampeador,
  faisaldebouni, fanizzamarco, gadial, gruu, hhorii, hykavitha, jagunther,
  jliu45, kanejess, klinvill, kurarrr, lerongil, ma5x, merav aharoni,
  michelle4654, ordmoj, rmoyard, saswati qiskit, sethmerkel, strickroman,
  sumitpuri, tigerjack, toural, vvilpas, welien, willhbang, yang.luh,
  yotamvakninibm, and {\v{C}}epulkovskis]{Qiskit}
Abraham,~H. \latin{et~al.}  Qiskit: An Open-source Framework for Quantum
  Computing. 2019\relax
\mciteBstWouldAddEndPuncttrue
\mciteSetBstMidEndSepPunct{\mcitedefaultmidpunct}
{\mcitedefaultendpunct}{\mcitedefaultseppunct}\relax
\EndOfBibitem
\bibitem[Li \latin{et~al.}(2013)Li, Yu, and Fei]{li2013geometry}
Li,~B.; Yu,~Z.-H.; Fei,~S.-M. Geometry of quantum computation with qutrits.
  \emph{Scientific reports} \textbf{2013}, \emph{3}, 1--6\relax
\mciteBstWouldAddEndPuncttrue
\mciteSetBstMidEndSepPunct{\mcitedefaultmidpunct}
{\mcitedefaultendpunct}{\mcitedefaultseppunct}\relax
\EndOfBibitem
\bibitem[Gardill \latin{et~al.}(2020)Gardill, Cambria, and
  Kolkowitz]{gardill2020fast}
Gardill,~A.; Cambria,~M.~C.; Kolkowitz,~S. Fast relaxation on qutrit
  transitions of nitrogen-vacancy centers in nanodiamonds. \emph{Physical
  Review Applied} \textbf{2020}, \emph{13}, 034010\relax
\mciteBstWouldAddEndPuncttrue
\mciteSetBstMidEndSepPunct{\mcitedefaultmidpunct}
{\mcitedefaultendpunct}{\mcitedefaultseppunct}\relax
\EndOfBibitem
\bibitem[Lucarelli(2002)]{lucarelli2002chow}
Lucarelli,~D. Chow's theorem and universal holonomic quantum computation.
  \emph{Journal of Physics A: Mathematical and General} \textbf{2002},
  \emph{35}, 5107\relax
\mciteBstWouldAddEndPuncttrue
\mciteSetBstMidEndSepPunct{\mcitedefaultmidpunct}
{\mcitedefaultendpunct}{\mcitedefaultseppunct}\relax
\EndOfBibitem
\bibitem[Iachello and Levine(1995)Iachello, and Levine]{iachello1995algebraic}
Iachello,~F.; Levine,~R.~D. \emph{Algebraic theory of molecules}; Oxford
  University Press, 1995\relax
\mciteBstWouldAddEndPuncttrue
\mciteSetBstMidEndSepPunct{\mcitedefaultmidpunct}
{\mcitedefaultendpunct}{\mcitedefaultseppunct}\relax
\EndOfBibitem
\bibitem[Staudemeyer and Morris(2019)Staudemeyer, and Morris]{LSTM}
Staudemeyer,~R.~C.; Morris,~E.~R. Understanding LSTM -- a tutorial into Long
  Short-Term Memory Recurrent Neural Networks. 2019;
  \url{https://arxiv.org/abs/1909.09586}\relax
\mciteBstWouldAddEndPuncttrue
\mciteSetBstMidEndSepPunct{\mcitedefaultmidpunct}
{\mcitedefaultendpunct}{\mcitedefaultseppunct}\relax
\EndOfBibitem
\bibitem[Sherstinsky(2020)]{Sherstinsky_2020}
Sherstinsky,~A. Fundamentals of Recurrent Neural Network ({RNN}) and Long
  Short-Term Memory ({LSTM}) network. \emph{Physica D: Nonlinear Phenomena}
  \textbf{2020}, \emph{404}, 132306\relax
\mciteBstWouldAddEndPuncttrue
\mciteSetBstMidEndSepPunct{\mcitedefaultmidpunct}
{\mcitedefaultendpunct}{\mcitedefaultseppunct}\relax
\EndOfBibitem
\bibitem[Schuld \latin{et~al.}(2019)Schuld, Bergholm, Gogolin, Izaac, and
  Killoran]{schuld2019evaluating}
Schuld,~M.; Bergholm,~V.; Gogolin,~C.; Izaac,~J.; Killoran,~N. Evaluating
  analytic gradients on quantum hardware. \emph{Physical Review A}
  \textbf{2019}, \emph{99}, 032331\relax
\mciteBstWouldAddEndPuncttrue
\mciteSetBstMidEndSepPunct{\mcitedefaultmidpunct}
{\mcitedefaultendpunct}{\mcitedefaultseppunct}\relax
\EndOfBibitem
\end{mcitethebibliography}

\end{document}